\documentclass[aps,prd,reprint,superscriptaddress]{revtex4-2}
\usepackage{slashed}
\usepackage{amsmath}
\usepackage{graphicx}
\usepackage{mathtools}
\usepackage[caption=false,position=top]{subfig}
\usepackage[export]{adjustbox}
\usepackage{dsfont}
\usepackage{xcolor}

\begin{document}

\title{Nonlinear effects in the excited states of many-fermion Einstein-Dirac solitons}


\author{Peter E.~D.~Leith}
\author{Chris A.~Hooley}
\author{Keith Horne}
\affiliation{SUPA, School of Physics and Astronomy, University of St Andrews, North Haugh, St Andrews, Fife KY16 9SS, UK}
\author{David G.~Dritschel}
\affiliation{School of Mathematics and Statistics, University of St Andrews, North Haugh, St Andrews, Fife KY16 9SS, UK}



\date{\today}

\begin{abstract}
We present an analysis of excited-state solutions for a gravitationally localized system consisting of a filled shell of high-angular-momentum fermions, using the Einstein-Dirac formalism introduced by Finster, Smoller, and Yau [Phys.~Rev.~D \textbf{59}, 104020 (1999)]. We show that, even when the particle number is relatively low ($N_f\ge 6$), the increased nonlinearity in the system causes a significant deviation in behavior from the two-fermion case.  Excited-state solutions can no longer be uniquely identified by the value of their central redshift, with this multiplicity producing distortions in the characteristic spiraling forms of the mass-radius relations. We discuss the connection between this effect and the internal structure of solutions in the relativistic regime.
\end{abstract}


\maketitle

\section{Introduction}
\label{secIntro}

The study of how quantum matter may interact within the framework of general relativity is an area in which much current research is focused. Without a fully satisfactory theory of quantum gravity, analysis of specific systems is usually limited to a semiclassical description, in which the gravitational field is treated as a purely classical object.

Of interest here is the study of gravitationally localized states, in which quantum particles are bound via their mutual gravitational attraction, but prevented from collapse by the effects of the uncertainty principle. We shall consider a system of $N_f$ neutral fermions, arranged in a spherically symmetric filled shell, localized solutions for which are found by solving the coupled Einstein-Dirac equations.

The existence of these structures relies heavily on the backreaction of the constituent matter on the spacetime metric, and as such they are difficult to model in a (perturbative) quantum field theory approach. The Einstein-Dirac formalism allows such solutions to be more readily analyzed, by treating the matter content as a first-quantized wavefunction, as opposed to a quantum field. The resulting solutions are therefore not fully quantum-mechanical, but may nonetheless provide a reasonable semi-classical approximation under which these objects may be studied.

Stable, gravitationally localized quantum states were first identified for bosonic systems, in the context of the coupled Einstein-Klein-Gordon system \cite{Kaup1968geon,Feinblum1968boson}, with the resulting objects ultimately becoming known as boson stars. Initial work on their fermionic counterparts was performed by Ruffini and Bonazzola \cite{Ruffini1969boson} and Lee and Pang \cite{Lee1987solitonStars}, but it was Finster \textit{et al.}\ \cite{FSY1999original} who provided the first numerically exact solutions, for the simplest case of two neutral fermions. These `particle-like', Planck-scale objects have been known variously as `fermion stars', `Dirac stars' and `Einstein-Dirac solitons'. It is the last of these which we shall adopt.

Subsequent work has extended this analysis to charged fermions \cite{FSY1999maxwell}, the addition of non-Abelian gauge fields \cite{FSY2000nonAbelianBound}, proofs of existence \cite{Bird2005proof,Nodari2010ED,Nodari2010EDM} and consideration of the Newtonian limit \cite{Stuart2010newtonianLimit}. Comparison between the fermionic and bosonic cases is presented in \cite{Blazquez2019bosonDirac} and \cite{Herdeiro2017bosonDiracProca}, while axisymmetric solutions corresponding to single fermion states have recently been found \cite{Herdeiro2019bosonDiracProcaSpinning}. Ground state solutions with large numbers of fermions have also been analyzed by the current authors \cite{Leith2020fermionTrapping}, and their structure interpreted in the form of a fermion self-trapping effect. In this paper, we shall consider the behavior of the corresponding excited states. 

Of particular relevance to this study is the work of Bakucz Can\'{a}rio \textit{et al.} \cite{Bakucz2020powerlaw}, who were able to find an exact solution to the massless Einstein-Dirac system, in which all metric and fermion fields scale as simple power laws. (We note that this solution, along with others, was independently found by Bl\'{a}zquez-Salcedo \textit{et al.}\ \cite{Blazquez2020ansatz}.) They proceeded to show the relevance of this exact solution to massive, high-redshift Einstein-Dirac solitons (relativistic states with a highly compressed central region), via a zonal classification of their internal structure. In what follows, we shall show that this classification requires alteration when considering systems with large numbers of fermions.

The paper is organized as follows. In Sec.~\ref{secEDsystem}, we formulate the equations of motion for a filled shell of $N_f$ fermions, using the Einstein-Dirac formalism. In Sec.~\ref{secGeneratingSolns}, we discuss how localized solutions to these equations can be generated numerically, and in Sec.~\ref{secN2} we review the known behavior of the two-fermion system. In Sec.~\ref{secVaryN}, we increase the particle number to $N_f=14$ and consider the effect on the first excited states, studying the development of the multi-valued regions that appear in the fermion energy curves, and the accompanying distortions of the mass-radius relations. In Sec.~\ref{secIndivSolns}, we extend to $N_f=38$, and analyze the resulting high-redshift solutions, showing that the increased nonlinearity in the system allows for additional variations in their internal structure. We move on to consider the behavior of higher excited states in Sec.~\ref{secHigherExcited}, before concluding with a discussion in Sec.~\ref{secConc}.

\section{Einstein-Dirac system}
\label{secEDsystem}

We begin with a brief outline of the derivation of the equations of motion for a filled shell of $N_f$ neutral fermions in the Einstein-Dirac formalism, a more detailed calculation of which can be found in \cite{FSY1999bhEDM}. We shall use the mostly positive metric convention $(-,+,+,+)$, and set $\hbar=c=1$. Note that factors of the Newton constant $G$ are retained in the following derivation, but when numerically generating solutions we shall set $G=1$.

The action for the Einstein-Dirac system can be written as
\begin{equation}
S_{\mathrm{ED}}=\int \left( \frac{1}{8 \pi G}R + \overline{\Psi}(\slashed{D}-m)\Psi \right )\sqrt{-g}\  \mathrm{d}^4x,
\label{EDaction}
\end{equation}
where $R$ is the Ricci scalar, $g=\mathrm{det}\left(g_{\mu\nu}\right)$, and $m$ is the fermion mass. Extremizing this action with respect to the spinor wavefunction $\Psi$ and metric $g_{\mu\nu}$ produces the Dirac and Einstein equations: 
\begin{align}
\left(\slashed{D}-m\right)\Psi&=0; \label{DiracEqn}\\
R_{\mu\nu}-\frac{1}{2}g_{\mu\nu}R&=8\pi G T_{\mu\nu}. \label{EinsteinEqn}
\end{align}

Using the vierbein formalism, the Dirac operator in curved spacetime can be written as $\slashed{D}=i\gamma^{\mu}\left(\partial_{\mu}+\Gamma_{\mu}\right)$, where $\Gamma_\mu$ is the spin connection and $\gamma^\mu$ are generalizations of the Dirac gamma matrices to curved spacetime, for which the anti-commutation relations $\left\{\gamma^{\mu},\gamma^{\nu}\right\}=-2g^{\mu\nu}$ hold.

We seek static, spherically symmetric solutions to these coupled equations, allowing the metric to be written, in the usual spherical co-ordinate system $(t,r,\theta,\phi)$, as
\begin{equation}
g_{\mu\nu}=\textup{diag}\left(-\frac{1}{T(r)^2},\frac{1}{A(r)},r^2,r^2 \sin^2 \theta\right),
\label{metric}
\end{equation}
where $T(r)$ and $A(r)$ are fields to be determined. With the metric written in such a way, a straightforward comparison with the Schwarzschild metric can be made, for which 
\begin{equation}
T_{\rm{Sch}}(r)^{-2}=A_{\rm{Sch}}(r)=1-\frac{2GM}{r},
\end{equation}
where $M$ would be the ADM mass of the localized state.

To allow for this simplification of spherical symmetry, the fermions must be arranged such that their total (spin + orbital) angular momentum is zero. One way to achieve this is to consider a filled shell, in analogy with an atomic orbital. In this case, the spinor wavefunction for each constituent fermion, having angular momentum $j$ with $z$-component $k$, can be written as
\begin{equation}
\Psi_{jk}=e^{-i\omega t}\frac{\sqrt{T(r)}}{r}\begin{pmatrix}\chi^k_{j-\frac{1}{2}}\alpha(r)\\i\chi^k_{j+\frac{1}{2}}\beta(r) \end{pmatrix}.
\label{fermionAnsatz}
\end{equation}
Each fermion oscillates at the same frequency $\omega$, ensuring that the overall wavefunction remains stationary. The fields $\alpha(r)$ and $\beta(r)$ are to be determined, while the two-component functions $\chi$ take the explicit forms
\begin{align}
\chi^k_{j-\frac{1}{2}}&=\sqrt{\frac{j+k}{2j}}Y^{k-\frac{1}{2}}_{j-\frac{1}{2}}\begin{pmatrix}1\\0\end{pmatrix}+\sqrt{\frac{j-k}{2j}}Y^{k+\frac{1}{2}}_{j-\frac{1}{2}}\binom{0}{1};\\
\chi^k_{j+\frac{1}{2}}&=\sqrt{\frac{j+1-k}{2j+2}}Y^{k-\frac{1}{2}}_{j+\frac{1}{2}}\begin{pmatrix}1\\0\end{pmatrix}\notag\\
&\hspace{80pt}-\sqrt{\frac{j+1+k}{2j+2}}Y^{k+\frac{1}{2}}_{j+\frac{1}{2}}\binom{0}{1},
\end{align}
where $Y^l_m(\theta,\phi)$ are the usual spherical harmonics. If desired, the total wavefunction can be reconstructed using the Hartree-Fock product,
\begin{equation}
\Psi=\Psi_{j,k=-j}\wedge\Psi_{j,k=-j+1}\wedge...\wedge\Psi_{j,k=j}.
\label{HartreeFock}
\end{equation}
For a filled shell with constituent fermions of angular momentum $j$, the number of fermions contained is therefore $N_f=2j+1$, where $j\in\{\frac{1}{2},\frac{3}{2},\frac{5}{2},...\}$. As such, we are limited to systems containing an even number of fermions. Note that the fermions are assumed to have positive parity (negative parity solutions do exist, but we shall not consider them here).

Upon restriction to a spherically symmetric filled shell, the following explicit form for the Dirac operator can be derived (see \cite{FSY1999original} for a detailed calculation):
\begin{multline}
\slashed{D}=i \gamma^t \frac{\partial}{\partial t}+i\gamma^r \left( \frac{\partial}{\partial r}+ \frac{1}{r}\left(1-\frac{1}{\sqrt{A}} \right) -\frac{T'}{2T}\right )\\+i\gamma^\theta \frac{\partial}{\partial\theta}+i\gamma^\phi \frac{\partial}{\partial\phi},
\end{multline}
where $'\equiv \mathrm{d}/\mathrm{d}r$. The curved space gamma matrices $\gamma^\mu$ are related to their flat space counterparts $\bar{\gamma}^a$ by the relation $\gamma^{\mu}=e^\mu_{\;\;a}\bar{\gamma}^a$, where in this case the only non-zero vierbein components are $e^{t}_{\;\;t}=T(r)$, $e^{r}_{\;\;r}=\sqrt{A(r)}$, and $e^{\theta}_{\;\;\theta}=e^{\phi}_{\;\;\phi}=1$.

Using this form for the Dirac operator, along with the metric (\ref{metric}) and fermion ansatz (\ref{fermionAnsatz}), the Dirac and Einstein equations reduce to the following set of four coupled differential equations for the four unknown fields $\alpha(r)$, $\beta(r)$, $A(r)$, and $T(r)$:
\begin{align}
\label{KappaEquations1}
\sqrt{A}\,\alpha'&=\frac{N_f}{2r}\alpha-(\omega T+m)\beta; \\
\label{KappaEquations2}
\sqrt{A}\,\beta'&=(\omega T-m)\alpha-\frac{N_f}{2r}\beta; \\
\label{KappaEquations3}
-1+A+rA'&=-8\pi GN_f\omega T^2(\alpha^2+\beta^2); \\
\label{KappaEquations4}
-1+A-2rA\frac{T'}{T}&=8\pi GN_fT\sqrt{A}\left(\alpha\beta'-\alpha'\beta\right).
\end{align}
This set of equations, along with appropriate boundary conditions, fully describes the gravitational interaction of a filled shell of fermions within the Einstein-Dirac formalism.

\section{Generating localized solutions}
\label{secGeneratingSolns}
We now discuss the method by which localized solutions to Eqs.~(\ref{KappaEquations1})--(\ref{KappaEquations4}) can be numerically generated. In terms of boundary conditions, the metric is required to be asymptotically flat, i.e.\ $T(r),A(r)\rightarrow 1$ as $r\rightarrow\infty$. In addition, we require the fermion wavefunction to be correctly normalized, giving the integral condition
\begin{equation}
4\pi\int_0^\infty(\alpha^2+\beta^2)\frac{T}{\sqrt{A}}\,\mathrm{d}r=1.
\label{eqNorm}
\end{equation}

Furthermore, we look for solutions that are regular (i.e.\ non-singular) at the origin, for which a unique asymptotic expansion exists, valid for small $r$:
\begin{align}
\label{kappaAsymptoticStart}
\alpha(r)&=\alpha_1r^{N_f/2}+...\;;\\
\beta(r)&=\frac{1}{N_f+1}(\omega T_0-m)\alpha_1r^{N_f/2+1}+...\; ;\\
T(r)&=T_0-4\pi GT_0^2\alpha_1^2\frac{1}{N_f+1}(2\omega T_0-m)r^{N_f}+...\; ; \\
A(r)&=1-8\pi G\omega T_0^2\alpha_1^2\frac{N_f}{N_f+1}r^{N_f}+...\; .
\label{kappaAsymptoticEnd}
\end{align}
This set of initial conditions adds two further parameters to the system --- $T_0$, the value of the metric field $T(r)$ at the origin, and $\alpha_1$, the initial slope of the fermion field $\alpha(r)$. 

Taking into account the conditions of asymptotic flatness and normalization is difficult from a computational point of view. We therefore make use of the rescaling technique introduced in \cite{FSY1999original} in order to convert these into a more manageable form. To do so, we temporarily set $T_0=m=1$ and look for solutions which instead obey the weaker conditions,
\begin{gather}
\tau=\lim_{r\rightarrow\infty}T(r)<\infty;\\
\lambda=4\pi\int_0^\infty \left(\alpha^2+\beta^2\right)\frac{T}{\sqrt{A}}\,\mathrm{d}r<\infty.
\end{gather}
These `unscaled' solutions are relatively straightforward to generate numerically. Upon choosing values for the remaining two unfixed parameters $\alpha_1$ and $\omega$, initial values for the fields are set using the small--$r$ expansion, and the numerical solver can proceed radially outwards. All that remains is to tune the value of $\omega$ such that the fermion fields $\alpha(r)$ and $\beta(r)$ tend to zero as $r\rightarrow\infty$. The true (physically relevant) solutions can then be obtained by rescaling the fields and parameters as follows:
\begin{align}
\alpha(r)&\rightarrow \sqrt{\frac{\tau}{\lambda}}\alpha(\lambda r);&
\beta(r)&\rightarrow \sqrt{\frac{\tau}{\lambda}}\beta(\lambda r);\notag\\
T(r)&\rightarrow \frac{1}{\tau} T(\lambda r);&
A(r)&\rightarrow A(\lambda r);\notag\\
m&\rightarrow \lambda m;&
\omega&\rightarrow \tau\lambda\omega.
\end{align}

When generating solutions using this method, the only parameter in the system that can be freely varied is the unscaled quantity $\alpha_1$. After the rescaling, however, the initial slope of $\alpha(r)$ cannot be used equivalently, and so we instead introduce a (physically relevant) parameter --- the central redshift $z\equiv T(0)-1$. This can take any value from $0$ to $\infty$, and, for the solutions presented here, is observed to be in one-to-one correspondence with $\alpha_1$. The central redshift can be interpreted as a measure of how relativistic a solution is, with $z\sim 1$ providing an approximate boundary between non-relativistic and relativistic cases.

Here, we generate solutions via the method outlined above using \textsc{Mathematica}'s built-in differential equation solver, NDSolve, with an explicit Runge-Kutta method. A one-parameter shooting method is implemented to determine the value(s) of $\omega$ for which the fermions become normalizable.
For the ground state solution, this takes the form of a simple binary chop, based on which axis is crossed in the $\alpha$--$\beta$ plane. A more involved technique is required, however, when generating excited states. As we shall show, for systems with $N_f\ge 6$, there is no longer always a unique solution for each excited state at a given value of redshift, and hence there may be more than one value of $\omega$ for which the fermions can be normalized. To ensure all solutions are found, we first perform an initial sweep of 500 $\omega$ values above the ground state, before focusing in on the regions that exhibit features indicating a solution may be present. Once all such regions are identified and isolated from each other, a binary chop can be used on each to determine the precise values of $\omega$.

Although this shooting method can be automated, it is nonetheless costly from a computational point of view, and there is an inevitable trade-off in terms of numerical precision. The precision required to generate solutions increases substantially with $N_f$, so our analysis is limited to systems with $N_f\lesssim 70$ fermions. Higher-redshift solutions also require a high precision to obtain, so our numerics in addition impose an upper limit in $z$. Nevertheless, these ranges are more than sufficient for a thorough analysis of the phenomena presented here.

\section{Review of $\boldsymbol{N_f=2}$}
\label{secN2}

\begin{figure*}
	\includegraphics{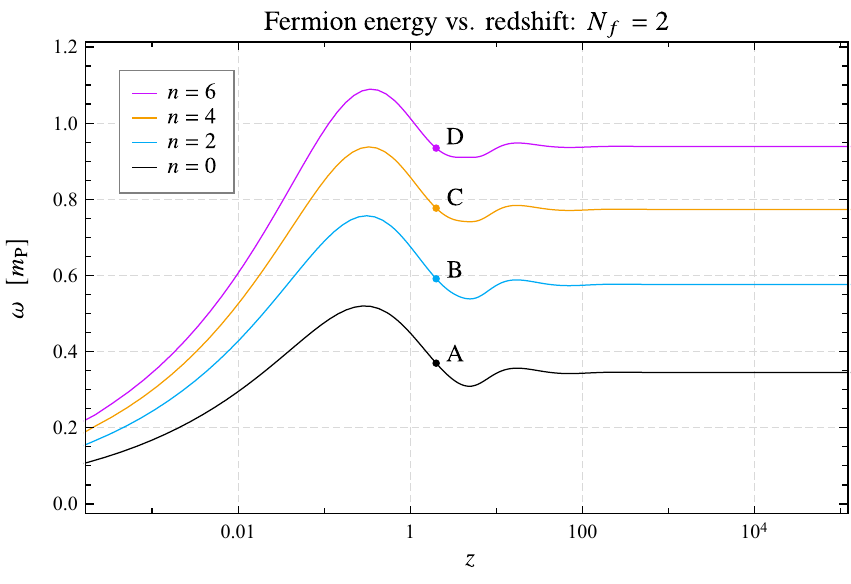}\hspace{10pt}
	\includegraphics{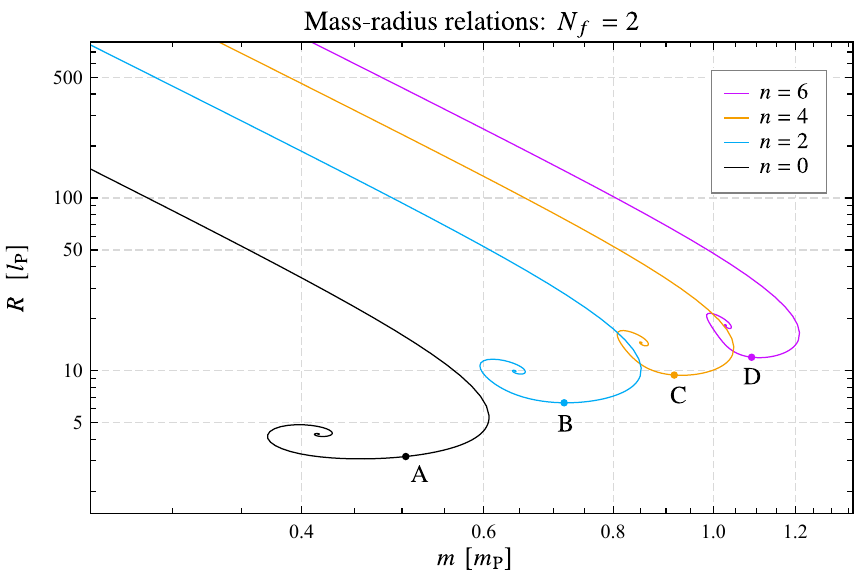}\\
	\vspace{5pt}
	\includegraphics{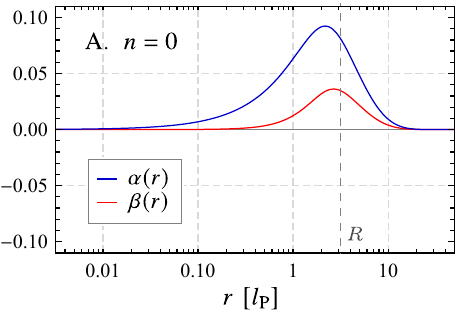}\hspace{5pt}
	\includegraphics{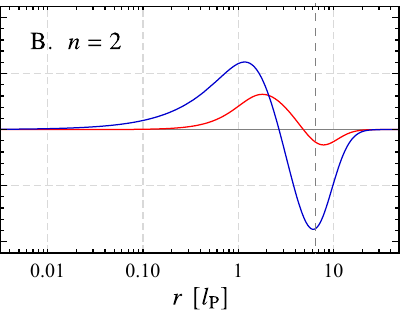}\hspace{5pt}
	\includegraphics{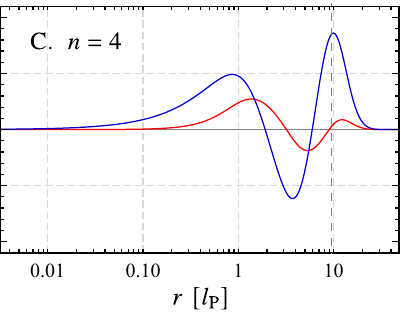}\hspace{5pt}
	\includegraphics{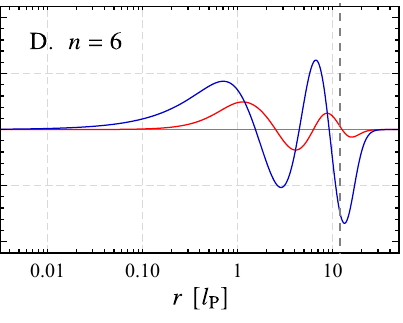}
	\caption{Properties of gravitationally localized states of two neutral fermions in the Einstein-Dirac approximation. \textbf{Top left}: The fermion energy $\omega$ as a function of the central redshift $z$, for the ground state and first three even-parity excited states. \textbf{Top right}: Mass-radius relations for the same four families of states. \textbf{Bottom row}: The radial structure of the fermion fields $\alpha(r)$ and $\beta(r)$ for the solutions marked A--D, each corresponding to a redshift value of $z=2$. The dashed line indicates the rms radius of the soliton. Note that all dimensionful quantities are measured in multiples of either the Planck mass, $m_\mathrm{P}=\sqrt{\hbar c/G}$, or the Planck length, $l_\mathrm{P}=\sqrt{\hbar G/c^3}$.}
	\label{figN2}
\end{figure*}

Before presenting results for large $N_f$, we first briefly review the known behavior of the two-fermion system initially studied by Finster \textit{et al.}\ \cite{FSY1999original}. For every value of the central redshift $z\in(0,\infty)$, there exists a unique ground state solution followed by an (infinite) series of excited states, distinguished by the number of zeros (nodes) in the fermion wavefunction. Note that we choose to label the $n^{\mathrm{th}}$ exited state as that in which the sum of the number of nodes in the fermion fields $\alpha(r)$ and $\beta(r)$ is equal to $n$. This labeling convention encompasses states of both positive parity ($n=2,4,6,8,...)$ and negative parity $(n=1,3,5,7,...)$. We shall not, however, consider negative parity states in what follows, since they form a separate branch of solutions, but we would expect them to exhibit a similar type of behavior to that shown here.

\begin{figure*}
	\includegraphics{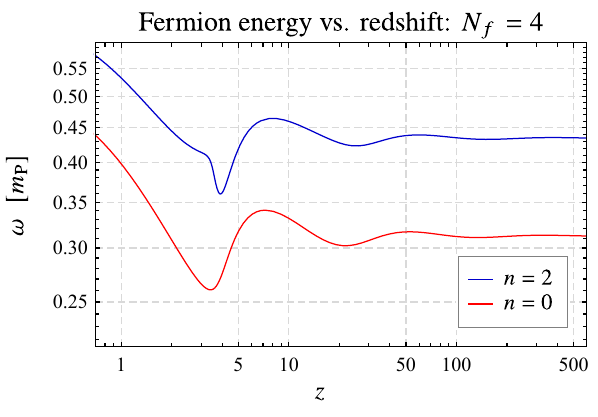}\hspace{5pt}
	\includegraphics{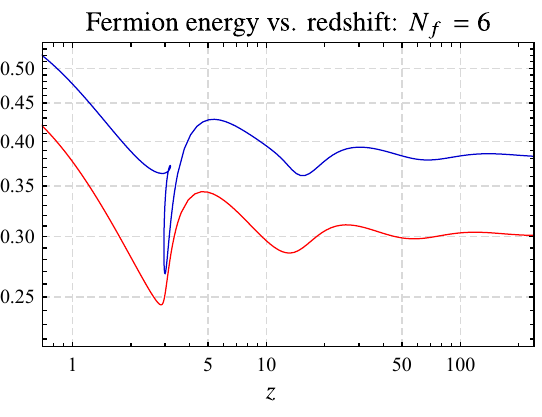}\hspace{5pt}
	\includegraphics{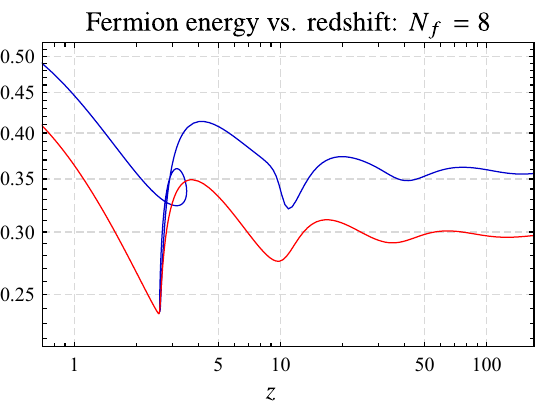}\\
	\vspace{3pt}
	\includegraphics{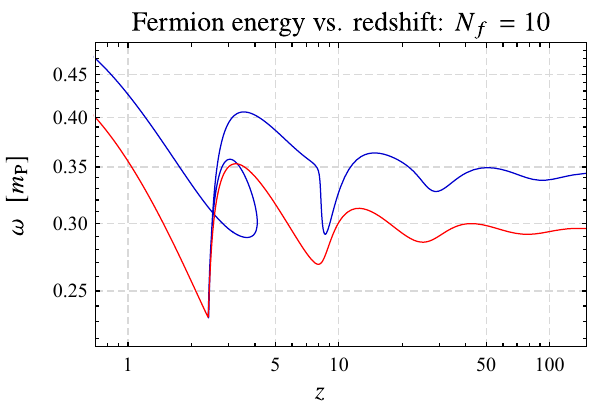}\hspace{5pt}
	\includegraphics{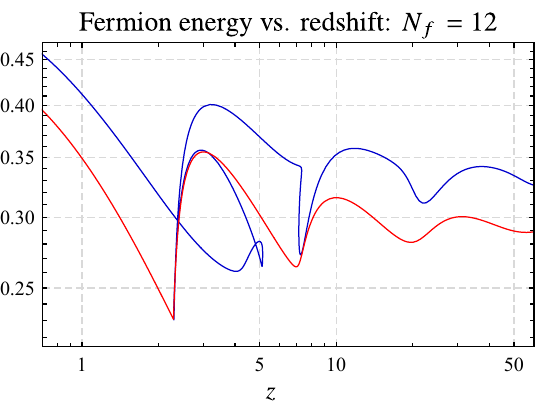}\hspace{5pt}
	\includegraphics{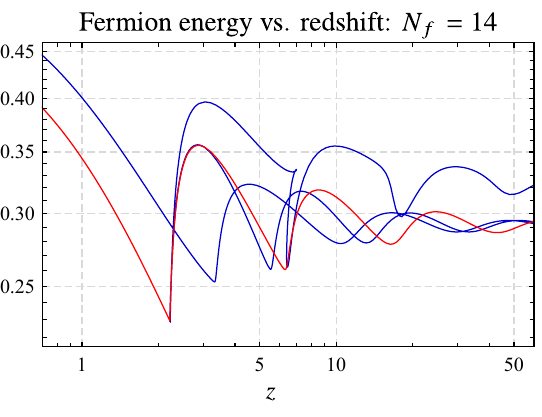}
	\caption{The fermion energy-redshift relations for the families of ground states ($n=0$) and first even-parity excited states ($n=2$) of many-fermion Einstein-Dirac solitons. These are plotted for six values of the fermion number, from $N_f=4$ to $N_f=14$. Note the development of a multi-valued region in the $n=2$ curve, the extent of which increases in redshift as $N_f$ is increased.}
	\label{figVaryNenergy}
\end{figure*}

The overall behavior of the two-fermion system is summarized in Fig.~\ref{figN2}. The top left panel shows how the fermion energy $\omega$ changes as the central redshift is varied, for the ground state and first three even-parity excited states. Each curve represents a continuous family of solutions parametrized by $z$, with the fermion energy increasing for each subsequent excited state. At low redshift, the curves approximate the expected non-relativistic relationship $\omega\propto z^{1/4}$ \cite{Leith2020fermionTrapping}, before the relativistic transition occurs at $z\sim 1$. This causes the onset of damped oscillatory behavior, with each curve oscillating around the appropriate infinite-redshift `power-law' solution \cite{Bakucz2020powerlaw}. 

The top right panel shows the mass-radius relations for the corresponding four families of states, where we define the radius $R$ of a soliton by 
\begin{equation}
R=\left(4\pi\int_0^\infty r^2\frac{T}{\sqrt{A}}\left(\alpha^2+\beta^2 \right )\mathrm{d}r \right )^{1/2},
\end{equation}
i.e.\ the rms radius weighted by the fermion density. It is worth emphasizing that the fermion mass is not a free parameter here --- its value is instead set by the rescaling procedure outlined in Sec.~\ref{secGeneratingSolns}. The mass-radius relations exhibit spiraling behavior, in common with models of astrophysical phenomena such as white dwarfs and neutron stars, and there exists a maximum mass, analogous to e.g.\ the Chandrasekhar limit. At low redshift, these curves approximate the non-relativistic relation $m\propto R^{-1/3}$ \cite{Leith2020fermionTrapping}, before spiraling towards their respective infinite-redshift solutions.

Marked on both these plots are four points, one located along each curve at a common redshift value of $z=2$. The individual solutions that occur at these points are shown in the bottom four panels, where we have plotted the radial structure of the fermion fields $\alpha(r)$ and $\beta(r)$ for each of the four states. In the ground state solution, both fermion fields are strictly positive, whereas an additional node arises in each field for each subsequent excited state. Note that, as mentioned, the $n^{\mathrm{th}}$ excited state contains a total of $n$ nodes in the fermion fields.

The internal radial structure of these states is not immediately obvious from these plots, but it has been demonstrated that there can exist up to four distinct zones within each solution \cite{Bakucz2020powerlaw}. The inner-most of these, referred to as the `core', is a region in which the fields roughly obey the small-$r$ asymptotic expansions (\ref{kappaAsymptoticStart})--(\ref{kappaAsymptoticEnd}). If the system is relativistic ($z\gtrsim 1$), the solution then transitions into a `power-law zone', in which the fields perform small-amplitude oscillations around the massless `power-law' solution. These damped oscillations, which are evenly spaced in $\log(r)$ with an envelope decreasing as $1/r$, are not strong
enough at $N_f=2$ to generate nodes in either of the fermion fields. Their number increases as $\log z$, while the radii at which they occur decrease as $1/z$. For excited states ($n>0$), a `wave zone' then follows containing the nodes that define the value of $n$. Finally, the solution enters the `evanescent zone', characterized by exponential decay of the fermion fields.

For relativistic solutions, it is also possible to separate the internal structure into sub-relativistic and relativistic regions based on the local value of the metric field $T(r)$. Since $T(r)$ monotonically decreases from its maximal central value, a single relativistic transition occurs at $T(r)\approx 2$, coinciding with the approximate end of the power-law zone. The relativistic region therefore encompasses the inner core and power-law zone, whereas the sub-relativistic region contains the wave and evanescent zones.

For the two-fermion case, the oscillations in the wave-zone are the dominant feature of the solutions, whereas the power-law oscillations are of too small amplitude even to be visible on the plots shown. For larger $N_f$, however, we shall see that this is no longer the case, and consequently the distinction between power-law and wave-zone oscillations is no longer as well-defined.

\section{Varying $\boldsymbol{N_f}$: 1st excited states}
\label{secVaryN}

We now present results showing how the behavior of excited-state solutions changes as we vary the number of fermions in the system. For $N_f\ge 6$, we shall find that multiple solutions can occur for the same value of the central redshift, and that this multiplicity has a significant effect on the structure of the mass-radius spirals. In this section and the next, we shall restrict our analysis to the first even-parity excited states ($n=2$), before moving to higher excited states in Sec.~\ref{secHigherExcited}.

\begin{figure*}
	\includegraphics{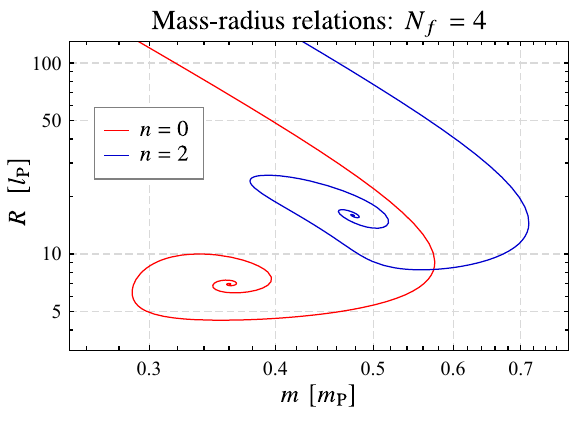}\hspace{5pt}
	\includegraphics{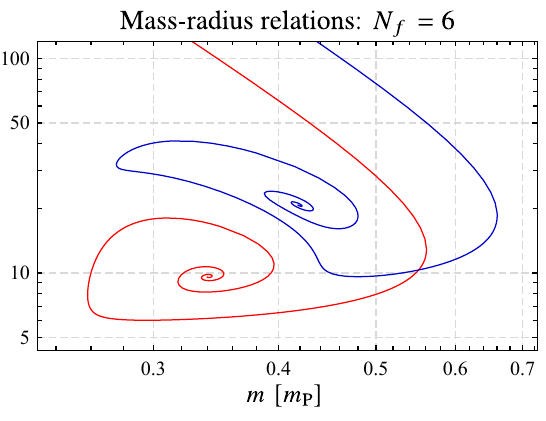}\hspace{5pt}
	\includegraphics{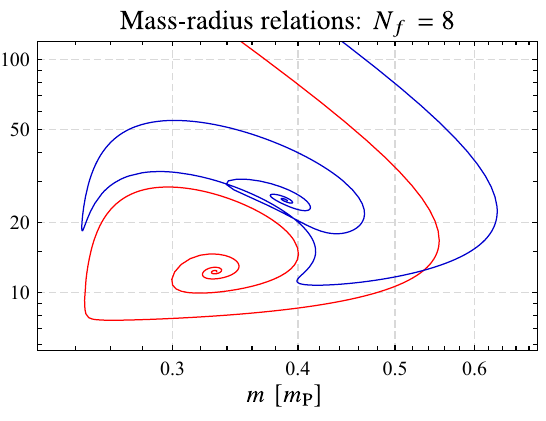}\\
	\vspace{3pt}
	\includegraphics{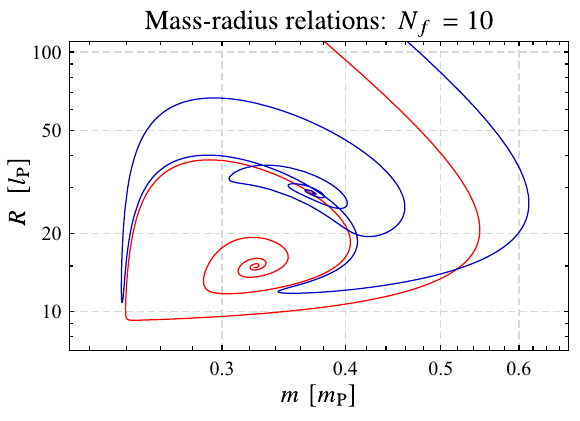}\hspace{5pt}
	\includegraphics{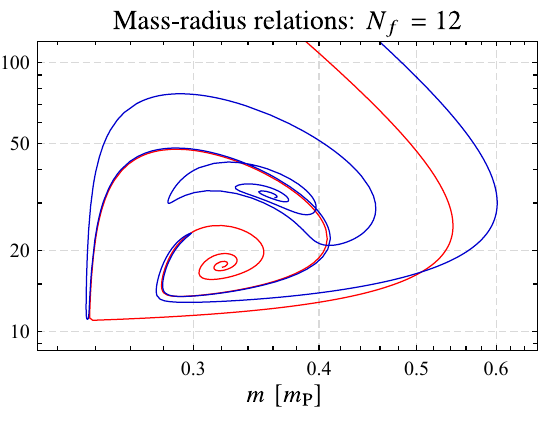}\hspace{5pt}
	\includegraphics{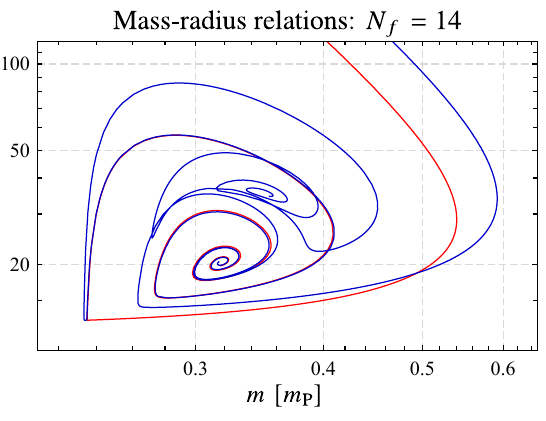}
	\caption{The mass-radius relations for the families of ground states ($n=0$) and first even-parity excited states ($n=2$), for the same set of $N_f$ values as in Fig.~\ref{figVaryNenergy}. For $N_f=6$, a visible distortion in the $n=2$ curve appears, and this region begins to wrap itself around the $n=0$ spiral as $N_f$ is increased. Once $N_f=14$ is reached, this effect has become so extreme that the $n=2$ curve appears to complete two spirals around both the $n=0$ and $n=2$ infinite-redshift solutions.}
	\label{figVaryNSpirals}
\end{figure*}

Fig.~\ref{figVaryNenergy} illustrates how the fermion energies of the families of ground and $n=2$ states change as the fermion number is increased from $N_f=4$ to $N_f=14$. We have isolated the relativistic portions of these curves, since prior to this there are no significant differences from the two-fermion case. First note the behavior of the ground state curve: as $N_f$ is increased, the amplitude of the oscillations becomes larger, and the first few minima develop into sharp points. We have previously interpreted this behavior by way of a fermion self-trapping effect (see \cite{Leith2020fermionTrapping} for details), where these sharp points correspond to the sudden appearance of new trapping regions in the solutions corresponding to those redshift values.

The $n=2$ curve, however, exhibits some additional, unexpected behavior. As $N_f$ is increased, a distortion develops, visible even for $N_f=4$, in the region surrounding the minimum of the first oscillation. From $N_f=6$ onwards, this distortion causes a portion of the curve to become multi-valued, while the locations of the first minima of the $n=0$ and $n=2$ curves gradually converge towards a common point. The overlapping region, or `fold', extends outwards in redshift as $N_f$ is increased, with the curve initially overshooting the minimum of the ground state curve, before turning back on itself, reaching this minimum and then proceeding as first expected. By $N_f=14$, the turning point of the $n=2$ curve has extended beyond the maximum redshift limit of our numerics, while the curve itself appears to temporarily oscillate around the ground state power-law solution.

In addition, notice that the second oscillation in the $n=2$ curve begins to exhibit the same behavior as the first, from $N_f=8$ onwards, approaching the second oscillation in the ground state curve. By $N_f=14$, this has developed into a further fold, and indeed the third oscillation displays the beginnings of a similar distortion. We therefore surmise that, if $N_f$ were to be increased further, each subsequent oscillation would ultimately behave in a similar fashion, with the curve becoming increasingly multi-valued in redshift as $N_f$ is increased. 

In spite of this multi-valued nature, it is important to emphasize that the curves nonetheless remain continuous for all values of $N_f$, i.e.\ they still represent a one-parameter family of solutions. The central redshift, however, is no longer the appropriate parameter to define that family. An attempt at obtaining a single-valued parameter for these curves is presented in the Appendix.

\begin{figure*}
	\includegraphics{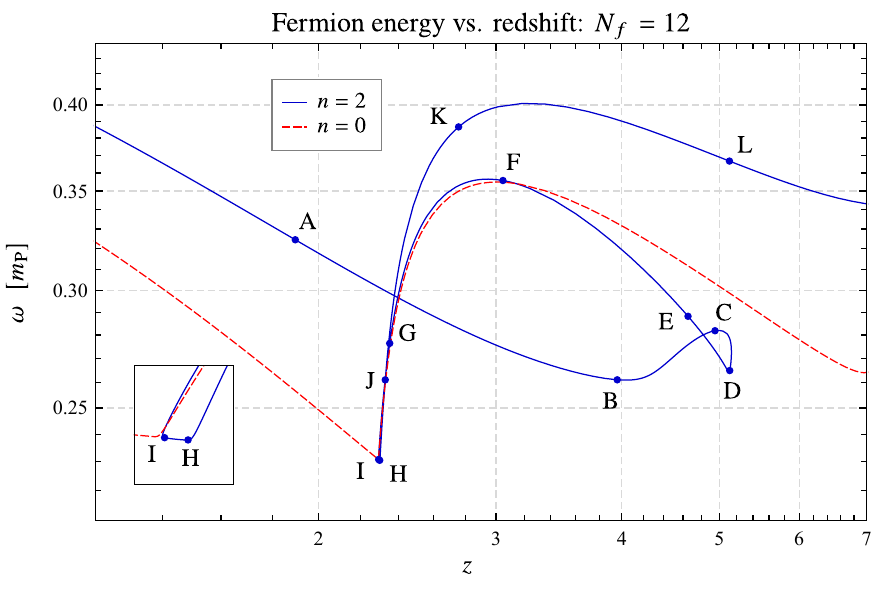}\hspace{5pt}
	\includegraphics{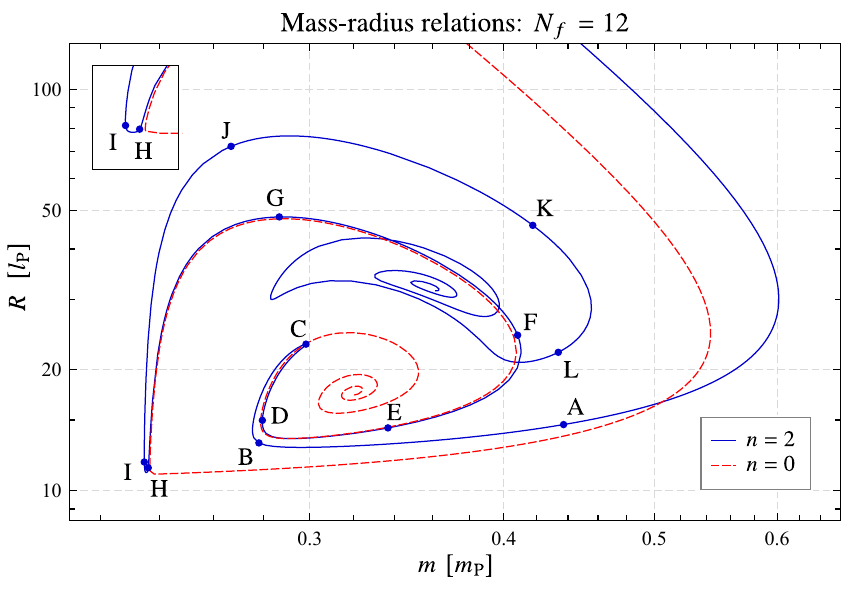}\\
	\vspace{5pt}
	\includegraphics{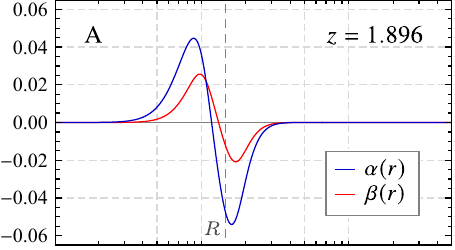}\hspace{5pt}	
	\includegraphics{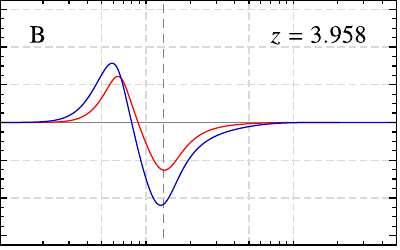}\hspace{5pt}
	\includegraphics{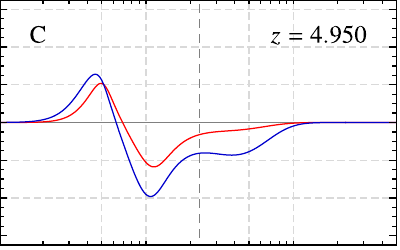}\hspace{5pt}
	\includegraphics{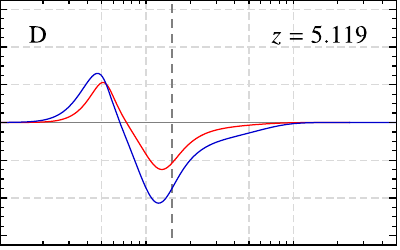}\vspace{5pt}
	
	\includegraphics{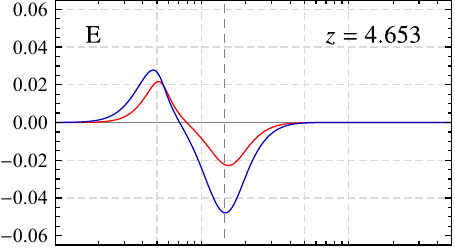}\hspace{5pt}	
	\includegraphics{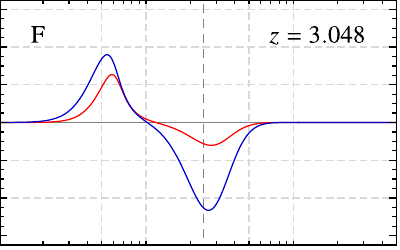}\hspace{5pt}
	\includegraphics{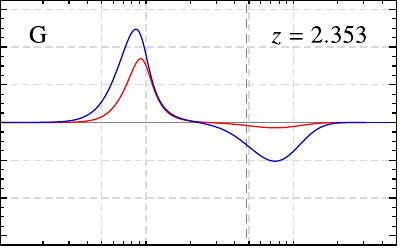}\hspace{5pt}
	\includegraphics{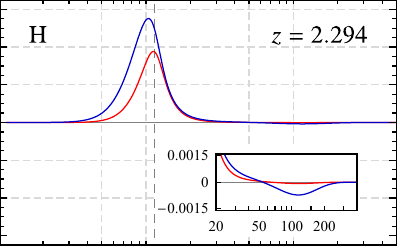}\vspace{5pt}
	
	\includegraphics{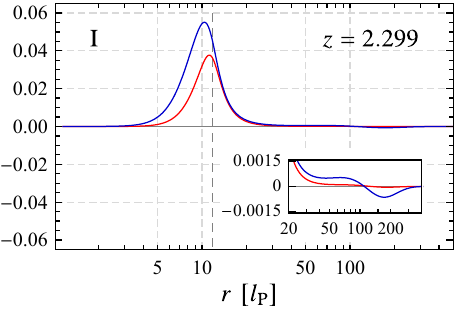}\hspace{5pt}	
	\includegraphics{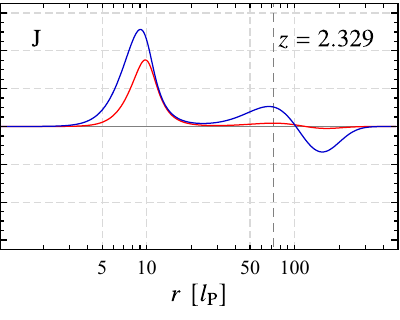}\hspace{5pt}
	\includegraphics{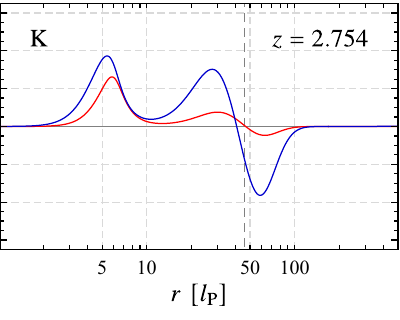}\hspace{5pt}
	\includegraphics{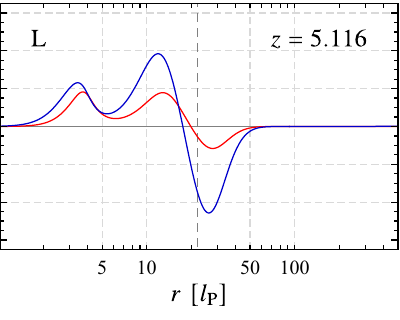}
	\caption{A study of the $N_f=12$ fermion system, showing the evolution in structure of the fermionic wavefunctions within the region of multi-valued redshift that occurs. \textbf{Top left:} The fermion energy-redshift relations of the ground and first even-parity excited-state families, focusing on this multi-valued region. \textbf{Top right:} The corresponding mass-radius relations for these families of states. \textbf{Bottom panels:} The radial structure of the fermion fields $\alpha(r)$ and $\beta(r)$ for twelve solutions located at the indicated points on the $n=2$ curves above. The redshift value of each is recorded in the upper right-hand corner of each plot.}
	\label{figN12loop}
\end{figure*}

The continuous nature of this family becomes more evident when considering the corresponding change in the mass-radius spirals, shown in Fig.~\ref{figVaryNSpirals}. Here, as $N_f$ is increased, a distortion appears in the $n=2$ spiral, and a portion of the curve begins to wrap itself around the ground state spiral. One end of this distortion ultimately becomes fixed near the first sharp turning point in the ground state curve, while the other moves progressively inwards towards the center of the ground state spiral. Indeed, by $N_f=14$, the $n=2$ curve appears to complete two separate spirals. It first closely follows the ground state spiral, then reverses before completing a second spiral towards the expected $n=2$ infinite-redshift solution. As with the fermion energy plots, we can see that this behavior starts to repeat a second time, with a similar distortion appearing further along the $n=2$ curve from $N_f=12$ onwards.

Note two important points. First, it is not clear whether the $n=2$ curve ever truly reaches the center of the ground state spiral. Beyond $N_f=14$, the end of the fold extends into a redshift regime which we cannot access, so it is not possible to ascertain whether the curve turns back at a finite value of redshift. This will be discussed further in Sec.~\ref{secConc}. Second, one might be concerned as to whether all solutions along these multi-valued curves should be classed as $n=2$ states, given that some can have a fermion energy lower than the ground state, or indeed have properties exceedingly close to those of a ground state solution. Given that the curves are still continuous, however, we feel that classifying states by counting the number of zeros in the fermion wavefunction continues to be the correct approach. We shall also continue to refer to states with $n>0$ as `excited', although here this does not necessarily imply higher fermion energy.

It is worth considering more closely the development of this multi-valued region, and in particular the evolution of the fermion wavefunction as we move along the curve. This is illustrated in Fig.~\ref{figN12loop} for the case of $N_f=12$, in which the two upper plots show the multi-valued portion of the fermion energy, along with the mass-radius relations. Indicated on the $n=2$ curves are the locations of the twelve solutions presented below, which show the radial structure of the fermion fields $\alpha(r)$ and $\beta(r)$ at various points along the curve.

The evolution of these individual solutions proceeds as follows. Solution A is located prior to entering the multi-valued region, and has the expected form for a non-relativistic $n=2$ state, with two extrema separated by a node in each fermion field. As we move into the multi-valued region, a second minimum, located outside the node, starts slowly developing, with its amplitude reaching a peak around solution C. The curve then reverses in redshift, however, and all trace of this minimum has disappeared entirely by solution E. As we move through solutions F and G, we can see that the amplitude of the fermion peak inside the node increases while that of the minimum outside the node decreases, and the region around the node deforms into an inflection point. We then approach the sharp turning point in the $n=2$ curve, either side of which are located solutions H and I (see zoomed regions). Solution H has the same structure as F and G, but now the amplitude of the minimum has decreased even further. Solution I looks ostensibly similar, but a zoom reveals that the inflection point has developed into a new maximum, located inside the node. The curve then proceeds forwards in redshift once more, with the amplitude of the two outer extrema increasing, until more recognizable $n=2$ states are obtained (K and L), these now containing two maxima at a smaller radius than the node.

\begin{figure*}
	\includegraphics{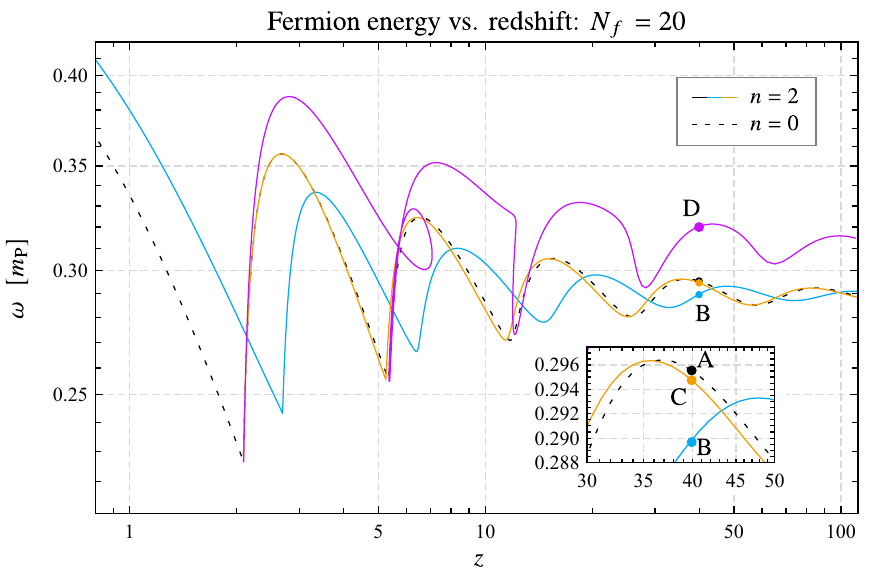}\hspace{5pt}
	\includegraphics{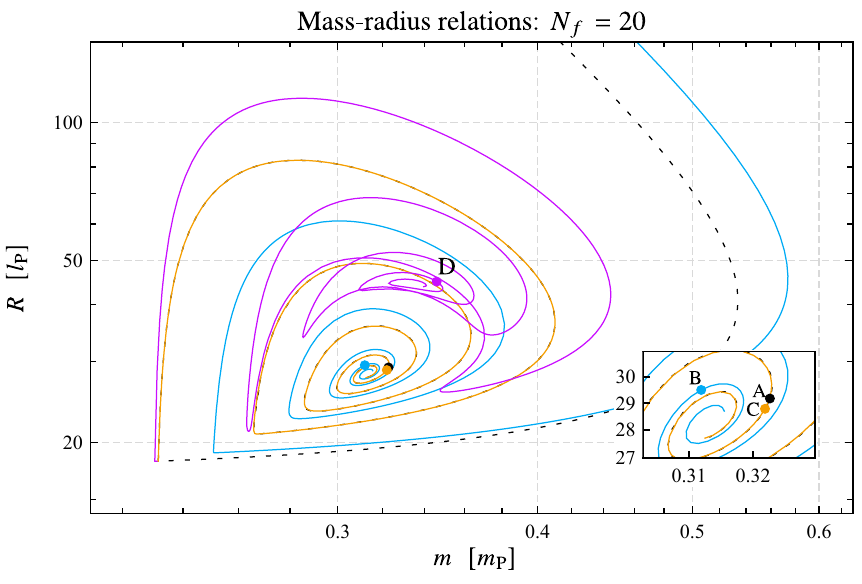}\\
	\vspace{5pt}
	\includegraphics{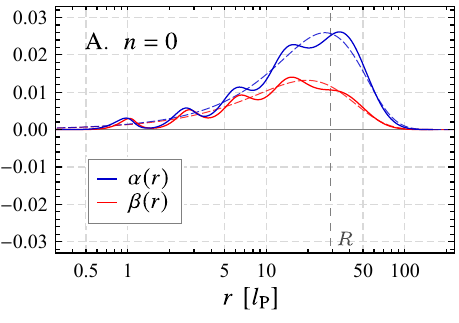}\hspace{5pt}
	\includegraphics{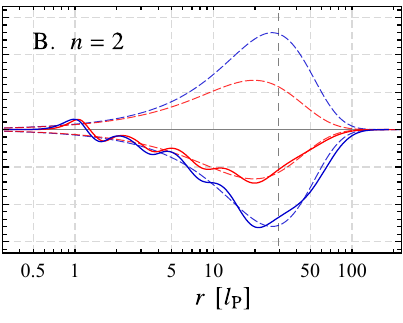}\hspace{5pt}
	\includegraphics{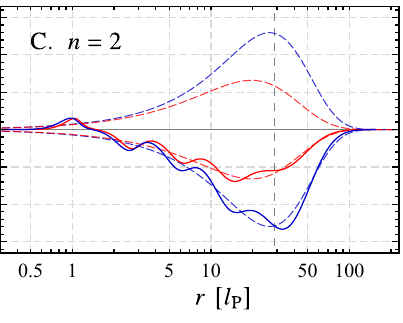}\hspace{5pt}
	\includegraphics{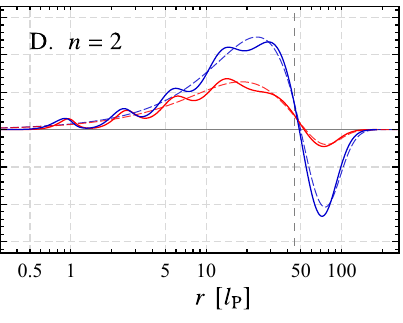}
	\caption{Plots summarizing the behavior of the $N_f=20$ fermion system. \textbf{Top left:} The fermion energy $\omega$ as a function of redshift for the ground state family ($n=0$) and first even-parity excited state family ($n=2$). We have separated the $n=2$ curve into three sections, indicated by the color-coding. \textbf{Top right:} The corresponding mass-radius relations for the same two families of states. \textbf{Bottom panels:} The fermion fields of the ground state solution and three $n=2$ solutions that occur at a redshift of $z=40$. The locations of these are indicated on the plots above. Included on each plot as dashed lines are the fermion field profiles of the infinite-redshift solution that contains the same number of nodes within the wave zone.}
	\label{figN20}
\end{figure*}

This behavior differs in one important respect from the redshift evolution of the two-fermion system. For $N_f=2$, we never observe nodes of the fermion fields in the relativistic power-law zone, only in the sub-relativistic wave zone. In solution C, however, we can see a new power-law oscillation beginning to form outside the radii of the nodes in the fermion fields. As we move along the curve, there is then a transition of these nodes from the inner relativistic power-law zone to the outer sub-relativistic wave zone. This will be discussed further in Sec.~\ref{secIndivSolns}.

It is important to emphasize that the evolution shown in Fig.~\ref{figN12loop} is continuous --- solution A can be smoothly transformed into solution L via intermediate states, all of which contain only a single zero in each fermion field. This informs our opinion that all points along the $n=2$ curve should indeed be classified as first (even-parity) excited states. The reason that the curve so closely approaches that of the ground state in some regions can be seen by considering solutions H and I. In both these cases, the extrema around the nodes are of very small amplitude, and so viewed on a large scale the solution resembles an $n=0$ state. It is not surprising, therefore, that these solutions have properties very similar to those of a ground state.

\section{High-redshift solutions}
\label{secIndivSolns}

What causes this multiplicity in redshift, and why does the first even-parity excited-state curve temporarily oscillate around that of the ground state at large $N_f$? A partial answer to these questions can be obtained by considering the structure of individual solutions at high central redshift. These highly relativistic solutions contain an extended power-law zone, allowing their structural differences to be more easily identified.

Fig.~\ref{figN20} shows the behavior of the family of first even-parity excited states for $N_f=20$. As with $N_f=14$, the multi-valued portion of the fermion energy curve extends beyond the redshift limit of our numerics, and a section oscillates around the $n=0$ infinite-redshift solution. The mass-radius relation similarly shows the curve spiraling towards the center of the $n=0$ spiral, before reversing and spiraling towards the $n=2$ infinite-redshift solution. For clarity, we have separated the curve color-wise into three separate branches --- the incoming section that spirals towards the infinite-redshift ground state solution (light blue), the intermediate section that then spirals outwards again, traversing backwards in redshift (orange), and finally the section beyond the redshift reversal at $z=2.1$ (purple). At a sufficiently high redshift value, there are three distinct $n=2$ states, one occurring along each of the three branches. For the case of $z=40$, the fermion fields for these solutions are shown in the bottom panels of Fig.~\ref{figN20}, together with those of the ground state. Included on each plot as dashed curves are the fermion field profiles corresponding to the infinite-redshift solution that lies at the center of the appropriate spiral.

Consider first the structure of the ground state (solution A). At this value of $z$, there is an extended relativistic power-law zone in which both $\alpha(r)$ and $\beta(r)$ oscillate around the infinite-redshift solution. In comparison with the two-fermion case, these oscillations are of much larger amplitude, so much so that the first minimum in the fermion fields is very close to zero. We have previously demonstrated \cite{Leith2020fermionTrapping} that these oscillations can be interpreted in terms of a fermion self-trapping effect, with the positions of the fermion field peaks corresponding to the locations of stable, null circular geodesics (photon spheres) in the soliton spacetime. The fermion field minima occur at the locations of the accompanying unstable photon spheres, from which the fermions are effectively repelled. This trapping effect becomes progressively stronger as $N_f$ is increased, since the additional mass results in a more severe distortion of the spacetime. This accounts for the increase in amplitude of the power-law oscillations. Finally, note that the trapping effect becomes progressively weaker moving outwards in radius, resulting in the value of the fermion fields at each subsequent minimum being higher than the previous.

Now consider solutions B--D. All three contain a single node in both $\alpha(r)$ and $\beta(r)$, and can therefore be classified as $n=2$ states. They differ, however, in terms of structure. Solution D exhibits the standard structure observed for $N_f=2$, with the node in each fermion field occurring outside the power-law zone, within the sub-relativistic wave zone. In solutions B and C, however, the nodes in the fermion fields occur within the power-law zone, just outside the first peak in the power-law oscillations. The reason that nodes can form within the power-law zone appears to be related to the fact that the minimum of the first power-law oscillation is close to zero. Note that, outside the node radii, the fermion fields switch to oscillating around the infinite-redshift solution in which both $\alpha(r)$ and $\beta(r)$ are negative.

Solutions B and C lack a wave zone, containing instead a direct transition from power-law to evanescent zone. Note that, although we have separated these two solutions onto different branches, there are no significant structural differences between them. The precise behavior around the fermion nodes differs, and there is an additional oscillation in solution C, but such features cannot be consistently used to distinguish between the solutions that occur along these two branches.

The positions of the three solutions on the $n=2$ curve are marked on the fermion energy and mass-radius plots. Solution D lies along the branch that spirals towards the $n=2$ infinite-redshift state, whereas B and C lie on the branches that follow the ground state curve. Why is it that solutions B and C appear to have properties so similar to $n=0$ states? To explain this, first note that the majority of the fermion mass is in fact located in the outer regions of the soliton. The properties of a solution are therefore primarily determined by the form of the wavefunction at large $r$. Note also that all physically observable quantities involve only bilinears of the fermion fields $\alpha(r)$ and $\beta(r)$. These two factors imply that neither a change in sign of the fermion fields, nor the presence of nodes deep within the power-law zone, should significantly affect a solution's properties. At high redshift, therefore, the properties of a solution are overwhelmingly determined simply by the number of fermion nodes within the outer wave zone. Since solutions B and C contain no nodes in the wave zone, their properties are very similar to those of a ground state solution.

\begin{figure}
	\includegraphics{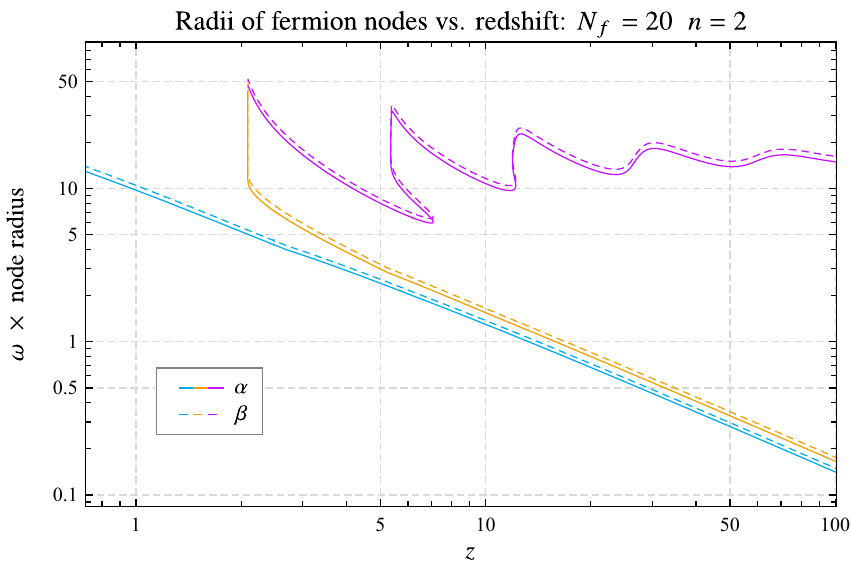}
	\caption{The change in radii of the fermion nodes as a function of central redshift, for the family of first even-parity excited states when $N_f=20$. The curve has been separated into the same three branches as in Fig.~\ref{figN20}. The effect of the fold is to carry the nodes radially inwards as redshift is increased. Note that along each branch the node in $\beta$ (dashed) always occurs at a larger radius than the node in $\alpha$ (solid).}
	\label{figN20nodes}
\end{figure}

Despite this, it is important to emphasize that solutions such as B and C should nonetheless be classified as $n=2$ states, since they contain a single node in each of the fermion fields. In addition, they are continuously connected to solution D via the $n=2$ curve, along which the fermion nodes transition from within the inner power-law zone to the outer wave zone. This transition is illustrated in Fig.~\ref{figN20nodes}, in which the radii of the fermion nodes are plotted as a function of redshift. Note that we have multiplied the node radii by a factor of $\omega$ in order to remove the oscillatory behavior that occurs within the power-law zone. The curves are color-coded in the same manner as Fig.~\ref{figN20}, with the light blue and orange portions assumed to meet at some (perhaps infinite) value of redshift beyond the maximum shown.

The evolution is as follows. The curve enters from the non-relativistic regime, and the nodes in the fermion fields move steadily inwards in radius as the redshift increases. This portion contains solutions in which the fermion nodes are located within the power-law zone, just outside the first peak in the power-law oscillations. Since the radial extent of the core shrinks as redshift is increased, the inner radius of the power-law zone begins becomes ever smaller, and the fermion nodes consequently move inwards. Once the curve reverses, the nodes move outwards again, occurring at slightly larger radii than those on the outgoing branch, albeit still within the power-law zone. This difference in radius can be seen in the structure of solutions B and C in Fig.~\ref{figN20}. The curve then gradually diverges from the lower branch as redshift decreases, until only a single power-law oscillation remains in the constituent solutions. This then allows the nodes to transition from the power-law zone to the wave zone, culminating in the sharp radial increase at $z=2.1$. The mechanism by which this occurs is similar to that shown previously in Fig.~\ref{figN12loop}. Beyond this transition, the curve reverses in redshift once again, and proceeds to oscillate around a constant radius. The solutions along this branch are those in which the fermion nodes are located within the wave zone. Note that the final branch contains a secondary multi-valued portion, or `fold', which is beginning to develop in a similar manner to the first. This can be seen also in the fermion energy plot in Fig.~\ref{figN20}. 

\begin{figure*}
	\includegraphics{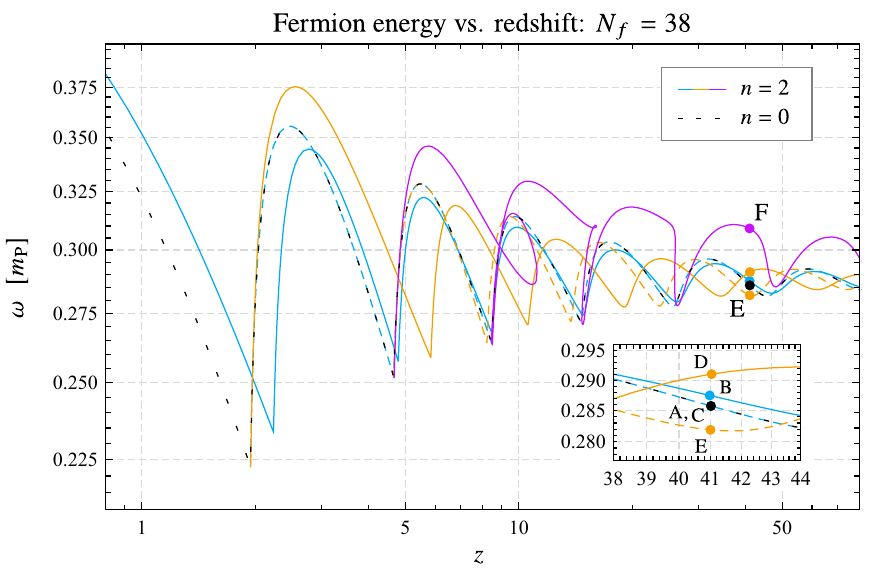}\hspace{5pt}
	\includegraphics{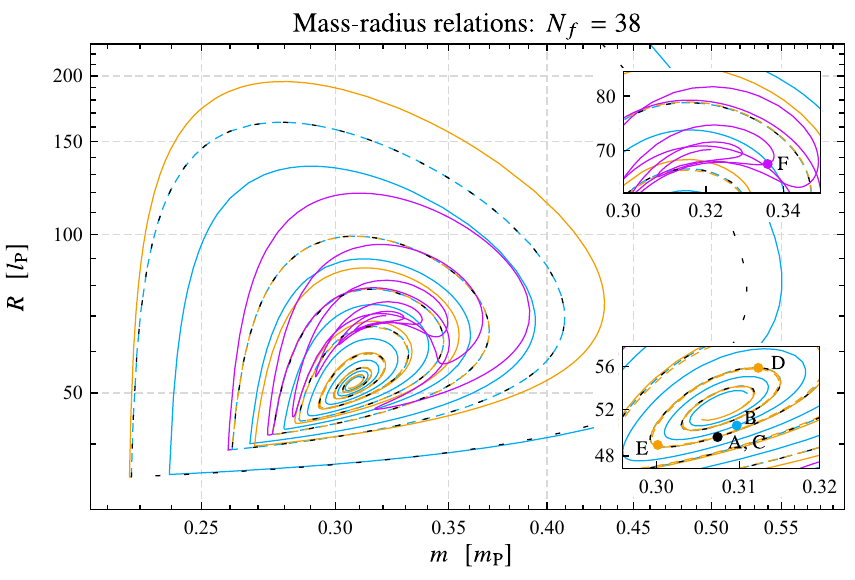}\\
	\vspace{5pt}
	\includegraphics{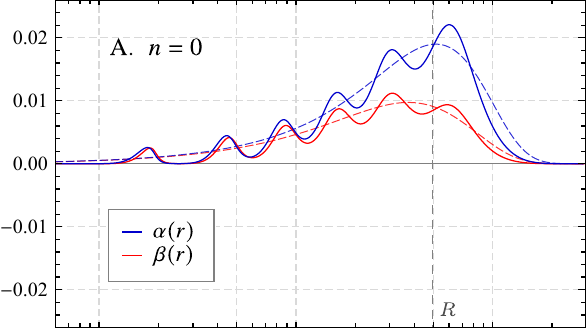}\hspace{8pt}
	\includegraphics{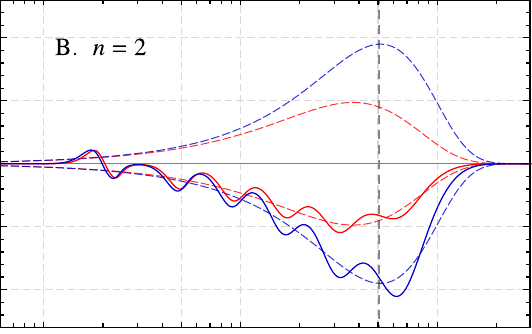}\hspace{8pt}
	\includegraphics{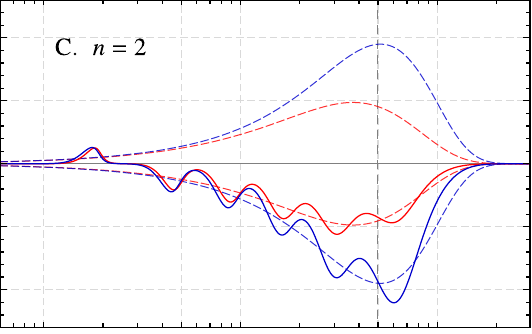}\\
	\vspace{8pt}
	\includegraphics{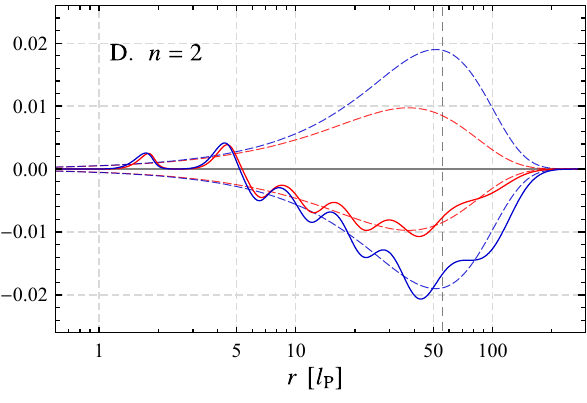}\hspace{8pt}
	\includegraphics{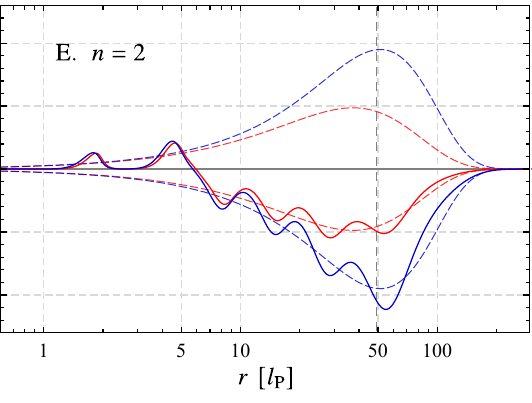}\hspace{8pt}
	\includegraphics{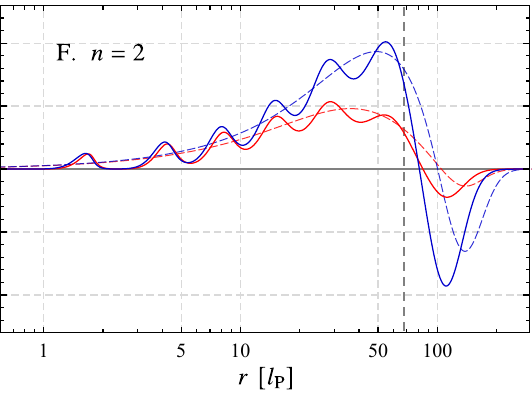}	
	\caption{A summary of the behavior of the $N_f=38$ fermion system. \textbf{Top left:} The fermion energy as a function of redshift for the ground and first even-parity excited-state families. There are now two folds in the $n=2$ curve, the first shown as light blue, and the second orange. The incoming branches of these folds are shown as dashed lines. \textbf{Top right:} The mass-radius plots for the $n=0$ and $n=2$ families. The $n=2$ curve spirals twice towards the $n=0$ infinite-redshift solution. \textbf{Bottom panels:} The fermion fields for the ground state solution and five $n=2$ solutions that occur at $z=41$, the locations of which are indicated on the fermion energy and mass-radius relations. Note that solutions A and C lie almost on top of each other.}
	\label{figN40n2}
\end{figure*}

In order to analyze the effects of this second fold, we must increase the value of $N_f$ until the fold extends into the high-redshift regime. To this end, Fig.~\ref{figN40n2} summarizes the behavior of the ground and first even-parity excited-state families for a system with $N_f=38$ fermions. We have again introduced a color-coding, whereby the $n=2$ curve transitions from light blue to orange to purple as we move continuously along it. The outgoing and incoming portions of each fold are now distinguished by solid and dashed lines, respectively. The behavior of the fermion energy and mass-radius relations is complicated, but can be briefly summarized as follows. Entering from the non-relativistic regime, the $n=2$ fermion energy curve oscillates, along the first fold (light blue), towards the $n=0$ infinite-redshift solution, before reversing and returning to $z=1.95$. Note that the incoming portion of this fold (dashed light blue) lies almost on top of the ground state curve. The curve then transitions to the orange branch, and oscillates for a second time around the $n=0$ infinite-redshift solution, once again reversing at a redshift value beyond the limit of our numerics. The final transition (from orange to purple) then occurs at $z=4.68$, after which the curve oscillates around the $n=2$ infinite-redshift solution. This behavior is mirrored in the mass-radius relation plot, in which the $n=2$ curve spirals twice towards the ground state infinite-redshift solution (along the light blue and orange branches), and once towards the $n=2$ infinite-redshift solution (along the purple branch).

It appears, therefore, that the effect of the second fold is similar to that of the first --- it also results in the $n=2$ curve temporarily spiraling towards the $n=0$ infinite-redshift solution. In order to ascertain the precise difference between these folds we must once again consider the structure of high-redshift solutions. The bottom panels of Fig.~\ref{figN40n2} show the fermion field profiles for the ground state and the five distinct $n=2$ states that now occur at $z=41$. Solutions B and C are located along the first fold, and have an identical nodal structure to their counterparts at $N_f=20$, with a single node in each fermion field appearing between the first and second power-law oscillations. Solution F, in which the nodes appear within the wave zone, is also present at $N_f=20$ (and indeed all $N_f$). The two new solutions, located along the second fold, are D and E, in which the fermion nodes now occur outside the second peak in the power-law oscillations. This new behavior is possible due to the increased strength of the second trapping region at high values of $N_f$, which lowers the minimum of the second power-law oscillation, allowing the fermion fields to change sign at this point. Note that, as is the case with the incoming and outgoing solutions along the first fold, there is no significant structural difference between solutions D and E. 

As discussed earlier, the overall properties of high-redshift solutions are primarily determined by the number of fermion nodes within the wave zone. Solutions B--E are therefore located near the center of the $n=0$ spiral, whereas F is on the branch that spirals towards the $n=2$ infinite-redshift solution. We once again emphasize, however, that all five solutions should correctly be classified as $n=2$ states, since they contain a single node in each fermion field, and are connected continuously by the $n=2$ curve. For $N_f=38$, there are now two important transition points along this curve. The first (from the light blue to the orange section) occurs at $z=1.95$, where the fermion nodes transition from just outside the first peak in the power-law oscillations to outside the second. The transition between the orange and purple branches then occurs at $z=4.68$, in which the nodes move outwards into the wave zone. This latter transition requires the presence of two power-law oscillations and thus takes place at a higher redshift than the first. Note that, although these points may appear sharp on the fermion energy and mass-radius plots, they are both in fact smooth continuous transitions.

We now have a reasonably complete picture for the behavior of the first even-parity excited states as we vary the number of fermions in the system. At small $N_f$, the power-law oscillations are of small amplitude, restricting the fermion nodes to the wave zone, and thus the fermion energy curve is single valued. As $N_f$ is increased, however, and the system becomes increasingly nonlinear, the first trapping region becomes strong enough that the first minimum in the power-law zone drops close to zero, allowing for the possibility of the fermion fields changing sign before the wave zone is reached. This results in a fold appearing in the $n=2$ fermion energy curve, containing new pairs of solutions in which the fermion nodes are located just between the first and second power-law oscillations. Along this fold, the mass-radius curve temporarily spirals towards the ground state infinite-redshift solution, since the solutions along it contain no nodes within the wave zone. At even higher $N_f$, the second trapping region becomes strong enough for a node to form between the second and third power-law oscillations. This creates another fold in the fermion energy curve, along with an accompanying new pair of solutions, and a second region that spirals around the ground state curve. We expect this pattern to continue as we increase $N_f$ further, with new solutions appearing in which the fermion nodes occur outside the third, fourth, and fifth peaks in the power-law oscillations, and so on. The fermion energy curve will therefore become increasingly multi-valued in redshift.

\section{Higher excited states}
\label{secHigherExcited}

\begin{figure*}
	\includegraphics{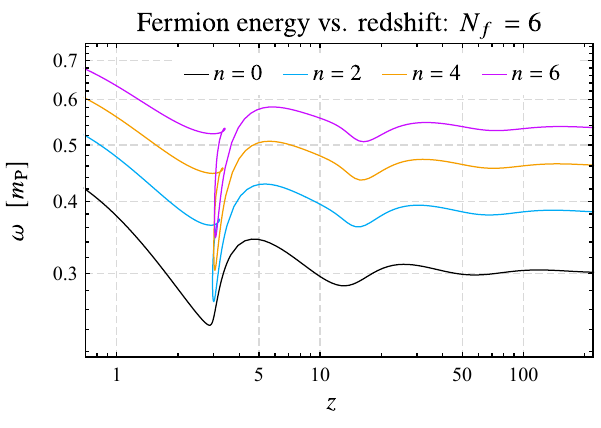}\hspace{5pt}
	\includegraphics{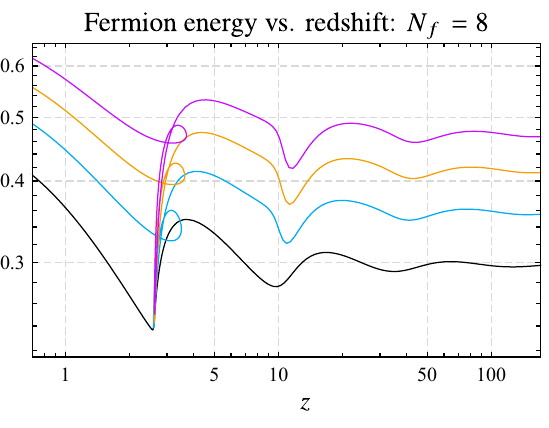}\hspace{5pt}
	\includegraphics{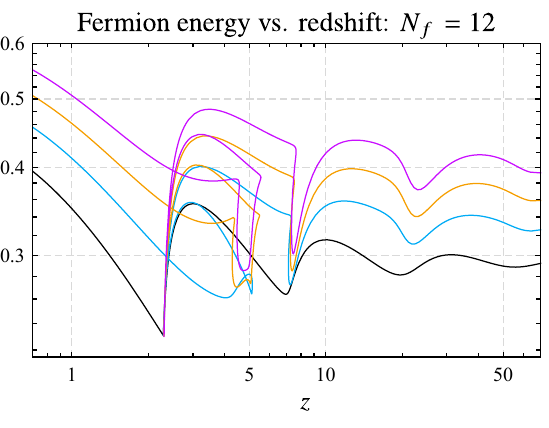}\\
	\vspace{3pt}
	\includegraphics{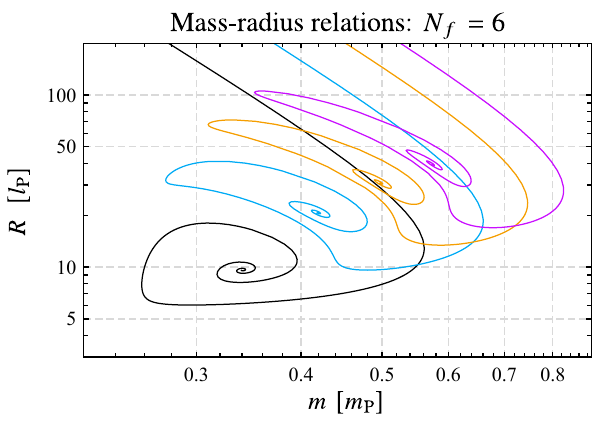}\hspace{3pt}
	\includegraphics{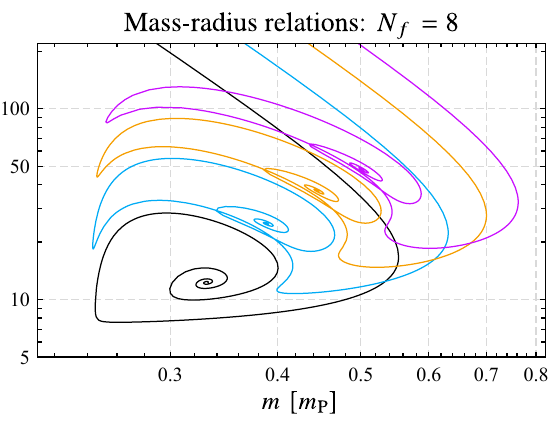}\hspace{3pt}
	\includegraphics{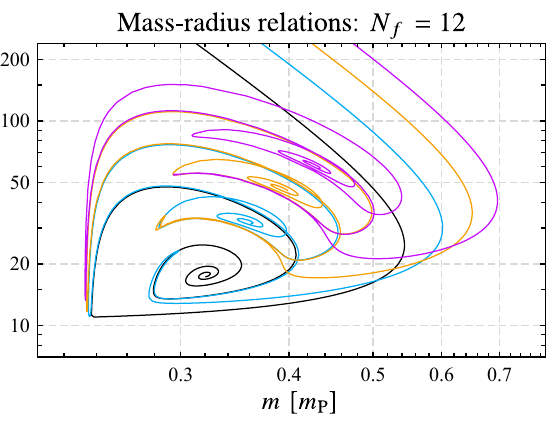}
	\caption{Plots showing how the ground and first three even-parity excited-state families behave for $N_f=6$, $N_f=8$ and $N_f=12$. \textbf{Top row:} The fermion energy-redshift relations for these four families of states, illustrating the parallel development of the first fold that appears in each of the three excited-state curves. \textbf{Bottom row:} The corresponding change in the mass-radius relations, in which each excited-state spiral begins to wrap around the curve directly below it.}
	\label{figExcitedVaryN}
\end{figure*}

So far, we have considered only the behavior of the first even-parity excited states ($n=2$). What happens to the higher excited states? In this section, we shall show that, at sufficiently large $N_f$, the number of solutions at constant redshift increases substantially with each subsequent family of states, owing to the increasing number of possible ways to distribute the (now multiple) fermion nodes within the power-law zone.

First, however, we shall detail how the higher excited states behave at relatively small $N_f$. This is illustrated in Fig.~\ref{figExcitedVaryN}, which shows the fermion energy and mass-radius relations for $N_f=6$, $N_f=8$ and $N_f=12$, for the ground state and first three even-parity excited-state families ($n=2,4,6$). These are to be compared with the two-fermion case shown in Fig.~\ref{figN2}. From the fermion energy plots, it is clear that the families of higher excited states behave in a similar manner to the first, with a fold appearing in each curve, moving to higher redshift as $N_f$ is increased. For $N_f=12$, the precise structure of the folds begins to differ, but they still extend over roughly the same redshift range. The mass-radius relations reveal a self-similar behavior, whereby each excited-state spiral begins to wrap around the curve directly below it as the folds develop. The additional solutions that arise due to this folding are also alike in internal structure for all excited-state families. Each fermion field contains a single node within the power-law zone, with the remainder located in the outer (wave) zone. Since the properties of a solution are dictated by the number of nodes in the wave zone, this explains why a portion of each curve follows that of the previous excited state --- the solutions located along the fold have one fewer wave-zone node in each fermion field than those on the rest of the curve. Given this behavior, it follows that the number of solutions present at a particular value of redshift will be the same for each excited-state family. For $N_f=20$, for example, there are three $n=4$ states at high redshift, the structures of which are identical to those of the three $n=2$ states shown previously in Fig.~\ref{figN20}, but with an additional pair of fermion nodes within the wave zone.

\begin{figure*}
	\includegraphics{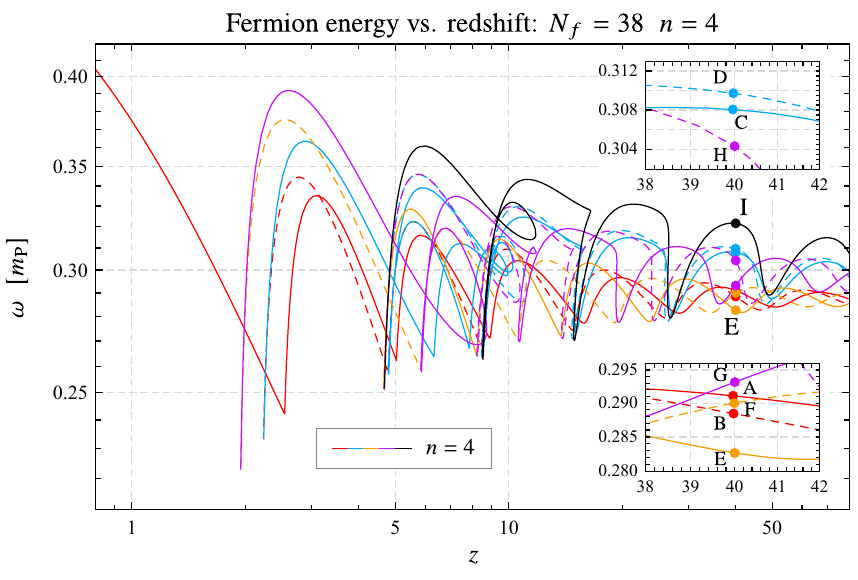}\hspace{5pt}
	\includegraphics{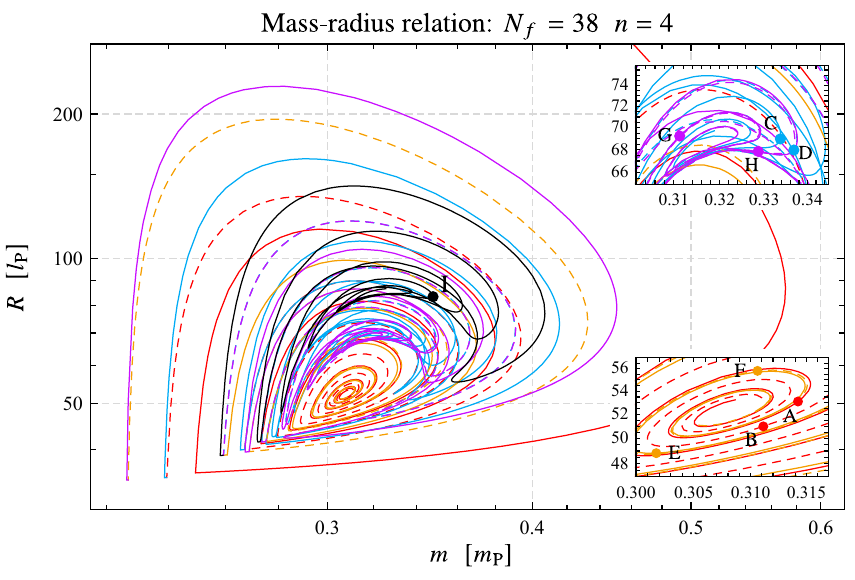}\\
	\vspace{5pt}
	\includegraphics{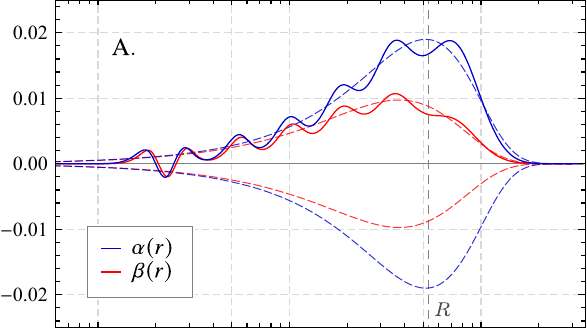}\hspace{8pt}
	\includegraphics{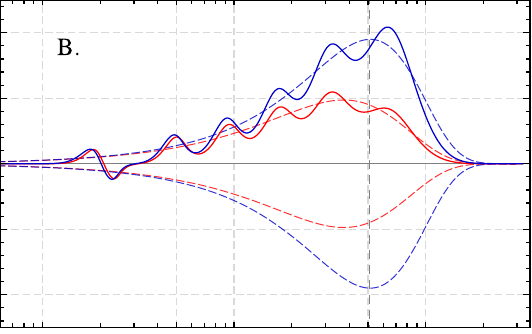}\hspace{8pt}
	\includegraphics{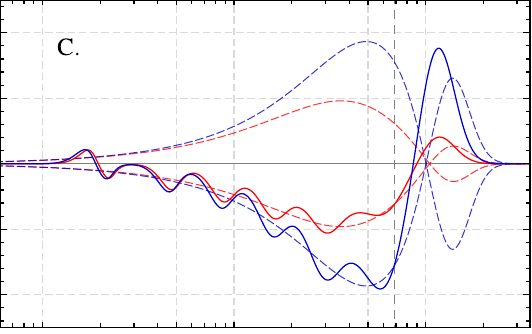}\\
	\vspace{8pt}
	\includegraphics{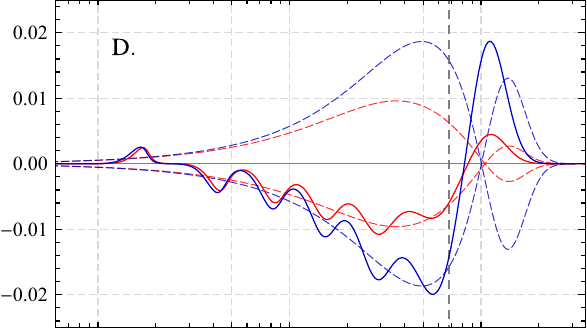}\hspace{8pt}
	\includegraphics{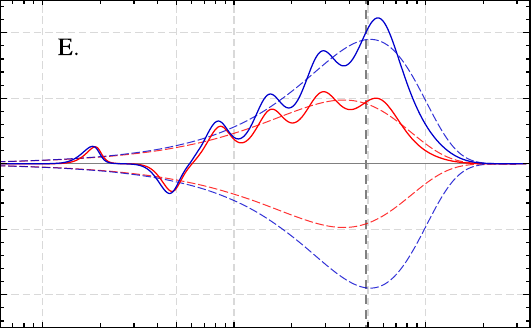}\hspace{8pt}
	\includegraphics{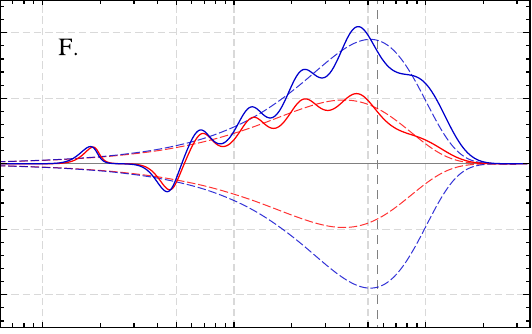}\\
	\vspace{8pt}
	\includegraphics{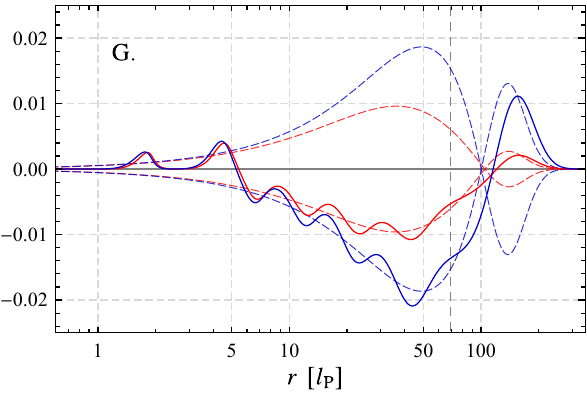}\hspace{8pt}
	\includegraphics{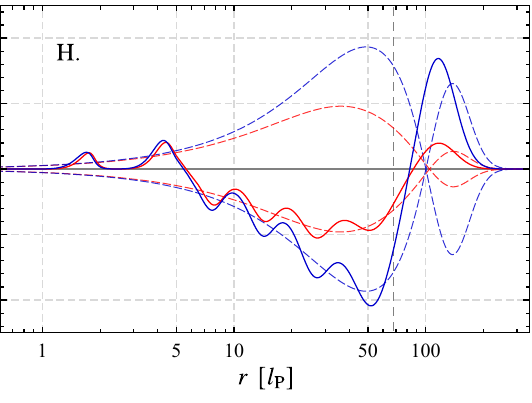}\hspace{8pt}
	\includegraphics{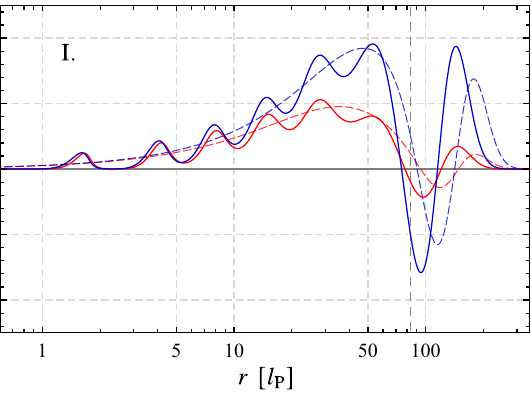}	
	\caption{Plots showing the behavior of the second even-parity excited state for $N_f=38$. \textbf{Top left:} The fermion energy curve for the family of $n=4$ states, now containing a total of four folds, each indicated by a different color. \textbf{Top right:} The mass-radius relation for the family of $n=4$ states, which spirals twice towards both the $n=0$ and $n=2$ infinite-redshift solutions. \textbf{Bottom panels:} The nine distinct $n=4$ solutions that occur at a redshift value of $z=40$. The locations of each solution along the mass-radius spiral can be predicted by counting the number of fermion nodes within the outer wave zone.
	}
	\label{figN40n4}
\end{figure*}

This landscape changes, however, as we increase $N_f$ further. Fig.~\ref{figN40n4} illustrates the behavior of the family of second even-parity excited states ($n=4$) for a system with $N_f=38$ fermions. Compared with the $n=2$ family (see Fig.~\ref{figN40n2}), the fermion energy curve contains two additional folds (making a total of four). We have again used color-coding to separate the curve into sections, transitioning from red $\rightarrow$ light blue $\rightarrow$ orange $\rightarrow$ purple $\rightarrow$ black as we move continuously along the curve. As before, the outgoing and incoming portions of each fold are represented respectively by solid and dashed lines. The evolution is complicated, but can be briefly summarized as follows. Entering from the non-relativistic (low redshift) regime, the fermion energy curve first oscillates around the $n=0$ infinite-redshift solution, before reversing at a redshift value beyond the maximum shown. It then transitions to the light blue branch at $z=2.24$, and moves forwards in redshift once again. It then oscillates around the $n=2$ infinite-redshift solution, reverses at high redshift, and then transitions to the orange branch at $z=1.95$. This behavior is then repeated, with the curve oscillating around that of the ground state along the orange branch, and around that of the $n=2$ excited state along the purple branch. The final transition occurs at $z=4.68$, after which the curve oscillates around the $n=4$ infinite-redshift solution. This behaviour is mirrored in the mass-radius relation, where it is clear that there are three distinct points towards which the $n=4$ curve spirals --- the $n=0$ infinite-redshift solution (along the red and orange branches), the $n=2$ infinite-redshift solution (along the light blue and purple branches), and the $n=4$ infinite-redshift solution (along the black branch). The primary difference in this behavior from that seen at lower $N_f$ is the fact that the $n=4$ curve spirals not just around the excited-state curve directly below it, but also around that of the ground state.

\begin{figure*}
	\includegraphics{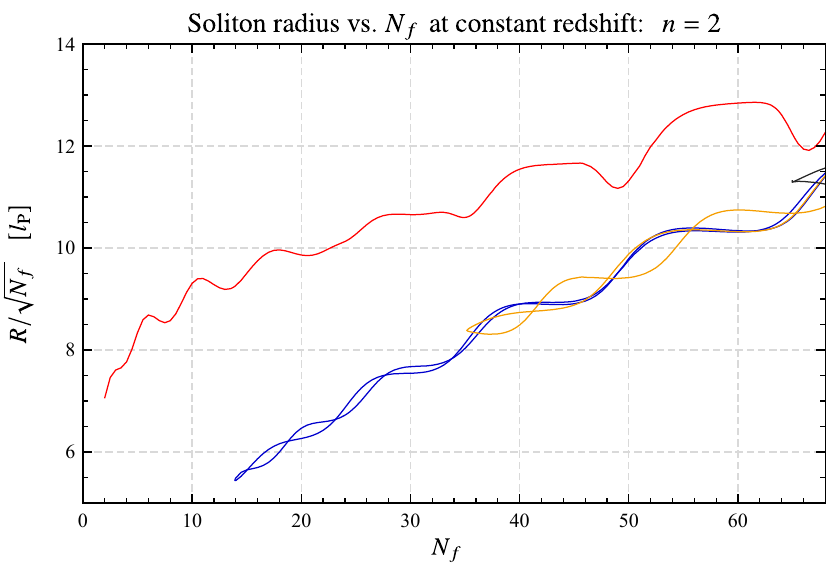}\hspace{10pt}
	\includegraphics{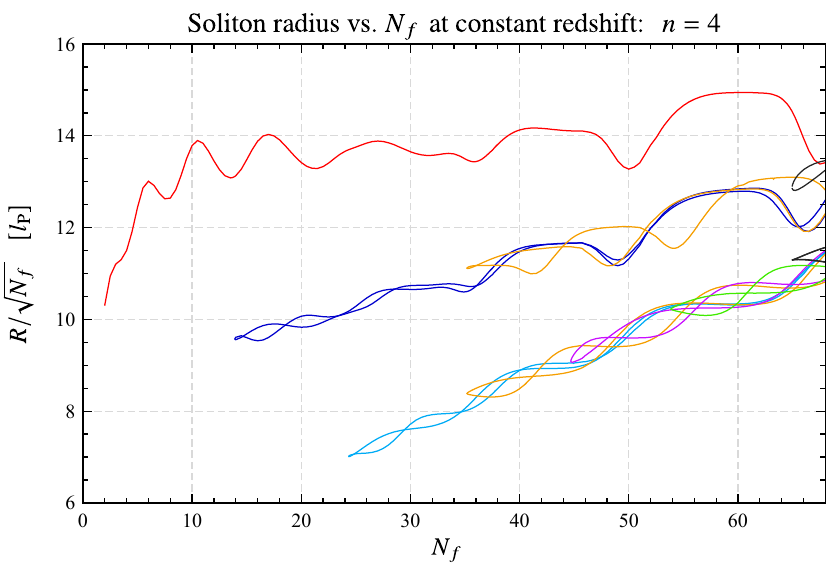}\\
	\vspace{5pt}
	\includegraphics{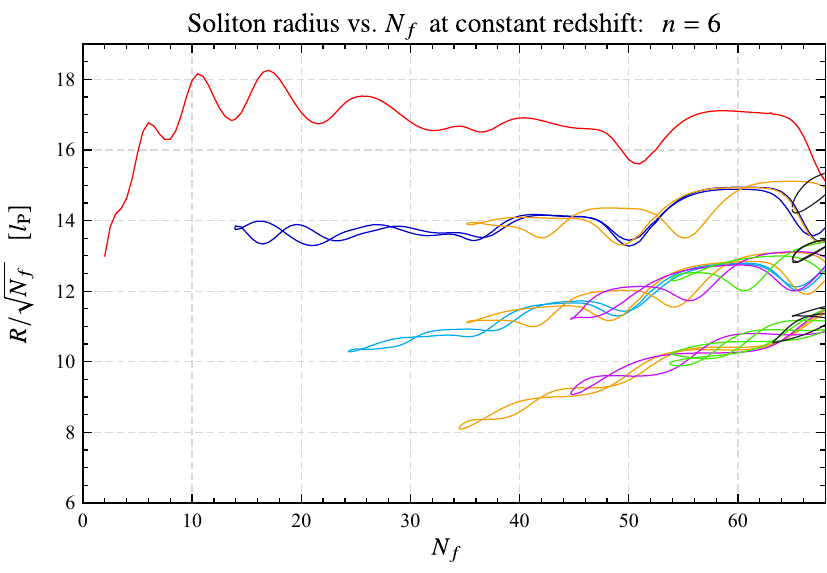}\hspace{10pt}
	\includegraphics{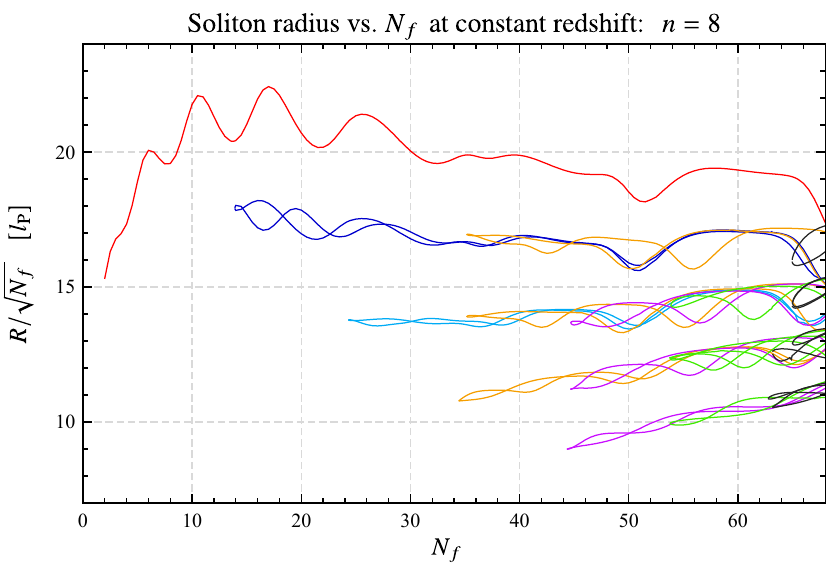}	
	\caption{Plots illustrating the behavior of the excited states of many-fermion Einstein-Dirac solitons, as the number of fermions in the system is varied. These show how the radii of the excited-state solutions with a redshift value of $z\approx 100$ change as a function of $N_f$, for $n=2$, 4, 6 and 8. Note that we included non-even and non-integer values of $N_f$ in this analysis. For systems with small numbers of fermions, only one of each excited state exists, but as $N_f$ is increased, new states emerge in pairs. We have color-coded the curves such that those that emerge at approximately the same $N_f$ are the same color.}
	\label{figNrad}
\end{figure*}

At high redshift, there are now nine distinct $n=4$ solutions, which are shown in the bottom panels of Fig.~\ref{figN40n4}, for the case of $z=40$. Five of these (solutions C, D, G, H and I) correspond to the $n=2$ states shown in Fig.~\ref{figN40n2}, with each containing an additional pair of fermion nodes within the outer (wave) zone. These are located along the branches of the curve that spiral towards the $n=2$ and $n=4$ infinite-redshift solutions. The four new solutions are A, B, E and F. In all four of these, both pairs of fermion nodes are located within the power-law zone, but it is not immediately obvious how to interpret the structural differences between those on the red branch (A and B) and those on the orange (E and F). We suggest that, in solutions A and B, both pairs of fermion nodes are located between the first and second power-law oscillations, whereas in solutions E and F, one is located outside the peak of the first oscillation and the other outside the peak (now a minimum) of the second. In all four cases, however, no nodes appear within the wave zone, and the solutions therefore lie along the branches that spiral towards the $n=0$ infinite-redshift solution.

Overall, we have shown that the complexity of the solutions increases substantially as we increase the number of fermions. This is further illustrated in Fig.~\ref{figNrad}, which summarizes the behavior of the first four even-parity excited states as a function of $N_f$. These plots show the radius of all excited-state solutions (for a given $n$) that are present at a constant redshift of $z\approx 100$. Strictly speaking, we have plotted $R/\sqrt{N_f}$ as this quantity varies only slightly as the fermion number is increased. Note that it is possible to solve Eqs.~(\ref{KappaEquations1})--(\ref{KappaEquations4}), along with the appropriate boundary conditions, for any strictly positive value of $N_f$. We have therefore included non-even and non-integer values of $N_f$ in this analysis, since it allows us to obtain a continuous picture of how the system varies. It is important to remember, however, that only systems where $N_f$ is an even integer correspond to physically realizable solitons.

Consider first the behavior of the first even-parity excited states (top left). For small fermion numbers, there is only a single $n=2$ state, in which the node in each fermion field is located within the wave zone. At $N_f=13.39$, however, a new pair of solutions emerges, originating from a single common point. These two states lie along the first fold in the fermion energy-redshift plots, and appear at the value of $N_f$ at which the end of this fold has first extended outwards in redshift to $z=100$. They correspond to solutions in which the fermion nodes are located within the power-law zone, just outside the first peak in the power-law oscillations, and hence have a significantly smaller radius than those with a wave-zone node. There remain only three solutions up until $N_f=35.19$, when a second pair of states appears, corresponding to the point when the second fold reaches $z=100$. These are the solutions in which the fermion nodes are located between the first and second peaks in the power-law oscillations. Finally, a third pair of states emerges at $N=64.98$, associated with the third fold, making a total of seven distinct solutions. If we were to increase $N_f$ further, we would expect each subsequent fold to result in the formation of an additional pair of solutions. Beyond some value of $N_f$, however, the redshift transition points from which new folds extend will begin to occur beyond $z=100$. Once this becomes the case, the number of states at $z=100$ will therefore remain constant, and observing new solutions would require moving to a higher value of redshift.

The equivalent behavior of the second even-parity excited states ($n=4$) is shown in the top right plot of Fig.~\ref{figNrad}. As for $n=2$, only a single solution exists for $N_f<13.39$, at which point the first new pair of states emerges, these containing one pair of nodes within the power-law zone and one within the wave zone. The behavior begins to differ as $N_f$ is increased, however, with a new pair of states appearing at $N_f=24.36$ that are not present at $n=2$. These are solutions with smaller radii, in which both nodes in each fermion field are located just outside the peak in the first power-law oscillations. At $N=35.19$, nodes can now form between the second and third power-law oscillations, with two new pairs of states emerging at this value, one in which all nodes are within the power-law zone, and the other in which one pair is located in the wave zone. Subsequent states then appear at $N_f$ values of 44.70, 53.88 and 64.98, making a total of 19 $n=4$ states at the point at which the limit of our numerics is reached.

The bottom two plots of Fig.~\ref{figNrad} show the behavior of the $n=6$ and $n=8$ states. Even more solutions are now present, due the increase in number of nodes, although the $N_f$ values at which new pairs emerge remain roughly the same as for $n=4$. The solutions congregate into distinct levels depending on their internal structure. The lowest level contains states in which no nodes exist within the wave zone, with the solutions on each subsequent higher level having one additional pair of wave-zone nodes. This allows us easily to read off the number of solutions, along with their overall structure, at a given value of $N_f$. For example, at $N_f=50$, there are 2 $n=8$ states with no wave-zone nodes, 4 with one pair of wave-zone nodes, 6 with two pairs, 4 with three pairs, and 1 with four pairs. The solutions within each level are distinguished by the precise distribution of the remaining nodes within the power-law zone.

\begin{figure}
	\includegraphics{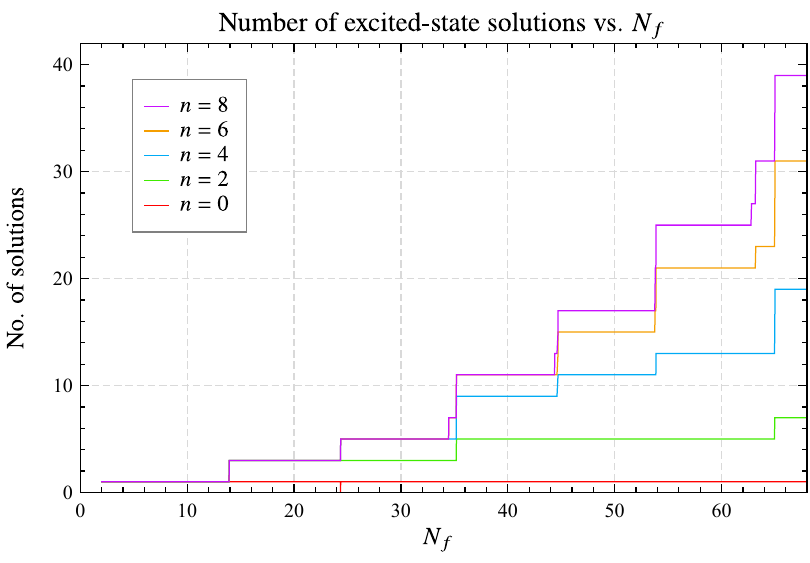}	
	\caption{A plot showing how the number of solutions for each of the first four families of even-parity excited-states (plus the ground state) changes as the number of fermions in the system is increased. The redshift value of each solution is roughly constant ($z\approx 100$). There remains a unique ground state in all cases, while the number of excited-state solutions steadily increases with $N_f$.}
	\label{figNoSolns}
\end{figure}

The change in the total number of excited-state solutions (again at $z\approx 100$) as $N_f$ is increased is summarized in Fig.~\ref{figNoSolns}. Below $N_f=13.93$, there is only a single solution for each family of excited states, and there are an equal number of solutions in each family until $N_f=24.36$. Beyond this, each higher excited-state family gradually gains additional solutions relative to the state family directly below, with this discrepancy growing ever larger as $N_f$ increases. By $N_f=68$ (the limit of our numerics), there are 7 distinct $n=2$ states, 19 $n=4$ states, 31 $n=6$ states and 39 $n=8$ states, along with a unique ground state. As mentioned previously, this difference arises because a larger number of nodes can be distributed in a larger number of ways within the power-law zone. This is amplified as $N_f$ is increased, since the stronger trapping effect results in more regions in which nodes can exist. Note that the number of solutions increases with roughly every 10 fermions added, and that the precise $N_f$ values at which this occurs differs slightly for each excited-state family.

\section{Discussion and Outlook}
\label{secConc}

We have shown that the behavior of the excited states of gravitationally localized states of many neutral fermions (Einstein-Dirac solitons) differs significantly from that of the two-fermion case, especially in the high-redshift (strongly relativistic) regime. Beyond $N_f=6$ fermions, the families of excited states are no longer single-valued in redshift, with a series of folds appearing in the fermion energy curves as we increase $N_f$. The appearance of these can be attributed to the existence of new, structurally distinct solutions in which one or more fermion nodes (zeros in the fermion wavefunction) are located within the relativistic power-law zone. Since a solution's properties are determined primarily by the number of nodes within the outer sub-relativistic wave zone, these folds cause the mass-radius relations of the higher excited-states to follow those of lower excited-states over a range of redshift values. The behavior of the system becomes increasingly complex as more fermions are added, with the number of excited-state solutions increasing at high redshift.

This picture is far from complete, however. We currently have very little physical understanding of this excited-state behavior, other than to remark that the system becomes increasingly nonlinear as we increase the number of fermions, and we should therefore not be surprised by the presence of multiple solutions. The increase in strength of the fermion self-trapping effect \cite{Leith2020fermionTrapping} (itself a result of the nonlinearity in the system) is certainly linked to the behavior, as it is ultimately responsible for allowing fermion nodes to appear within the power-law zone. This self-trapping explains why the minima in the fermion fields approach so close to zero, but it does not explain why these minima can transition into nodes. This could perhaps be partially understood in terms of an effective potential barrier that the fermion fields can overcome if the amplitude of the preceding power-law oscillation is sufficiently large. It is also unclear whether the precise location of the fermion nodes (e.g.\ within the power-law zone or wave zone) has any direct physical interpretation.

Although the self-trapping effect can provide some indication of which excited-state solutions may be present at a given $N_f$, we cannot precisely predict which solutions will occur at a particular value of redshift. One reason for this is that we do not currently understand what causes the folds in the fermion energy to end at the redshift values at which they do. Along each fold, the fermion nodes move inwards towards smaller radii, but at some point the curve reverses and the nodes move outwards again. Is there perhaps a mechanism that precludes the existence of nodes below a certain radius, the precise value of which depends on the value of $N_f$? For large enough $N_f$, it might be that the nodes are indeed able to reach $r=0$, with the fold therefore reversing only at strictly infinite redshift. If this is the case, then it may be that an analytic perturbation analysis at small $r$ could reveal a hidden degeneracy in the infinite-redshift solutions \cite{Bakucz2020powerlaw}.

In a broader context, we note that the excited-state behavior presented here may not be confined to Einstein--Dirac solitons. In particular, there does not appear to be a particularly strong reliance on the fermionic nature of the system, so we might expect to observe similar effects in objects such as boson stars. Although a shell of high-angular-momentum particles is not such a natural configuration in the context of bosons, comparison might be drawn with the case of rotating boson stars. Such objects have been widely studied, and indeed their excited states potentially show signs of a similar type of behavior to that discussed here \cite{bosonRotating}.

Finally, we briefly discuss the question of stability. Regardless of the value of $N_f$, we find that all solutions within the relativistic regime have a positive binding energy, and are therefore expected to be dynamically unstable. This does not mean, however, that the study of these solutions should be neglected --- unstable resonances are important in the field of particle physics, for example. It may also be possible to stabilize the solutions by coupling additional fields to the system. We shall consider the specific case of including a scalar Higgs field in a future publication. Regardless of the issue of stability, however, the results presented here nonetheless provide an interesting example of how the effects of nonlinearity can influence the interaction of quantum matter within the framework of general relativity.

\begin{acknowledgments}
P.~E.~D.~L.\ acknowledges funding from a St Leonards
scholarship from the University of St Andrews and from
UKRI under EPSRC Grant No.~EP/R513337/1.
\end{acknowledgments}

\appendix*

\section{A single-valued parametrization for the families of excited states}

In the analysis presented above, we have shown that, for many-fermion Einstein-Dirac solitons with $N_f\ge6$, excited-state solutions can no longer be uniquely identified by the value of their central redshift $z$. At a given value of redshift, multiple solutions can exist that belong to the same family of excited states, and consequently quantities such as the fermion energy become multi-valued when plotted as a function of $z$. As evidenced previously, however, the curves that define these families are still continuous, which implies that there must exist a single-valued parameter that increases monotonically as the curves are traversed. In what follows, we shall outline a possible method by which such a parameter can be obtained.

\begin{figure*}
	\includegraphics{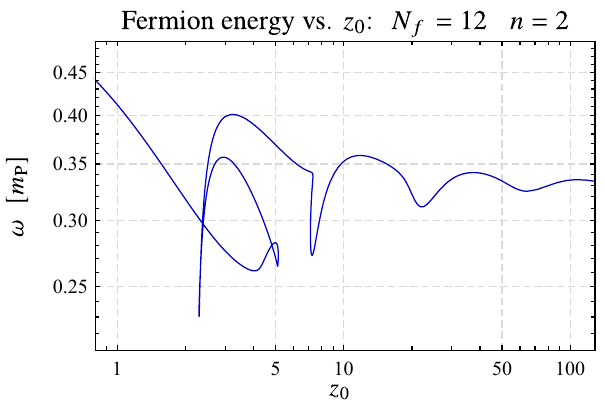}\hspace{5pt}
	\includegraphics{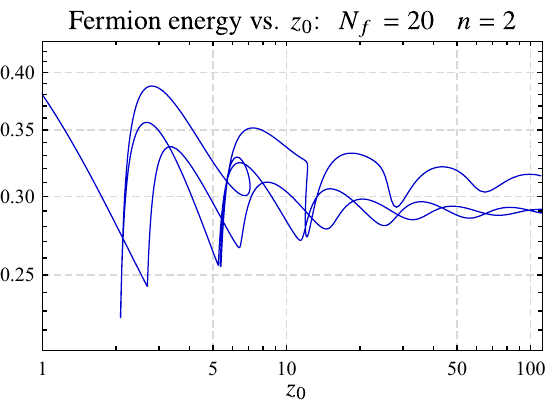}\hspace{5pt}
	\includegraphics{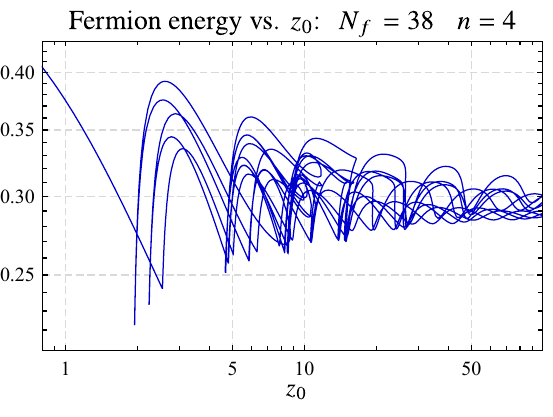}\\
	\vspace{5pt}
	\includegraphics{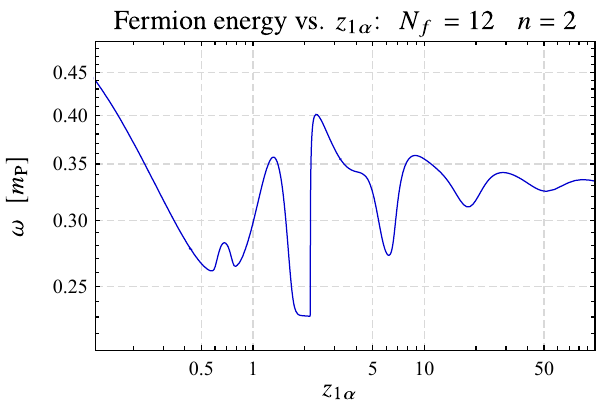}\hspace{5pt}
	\includegraphics{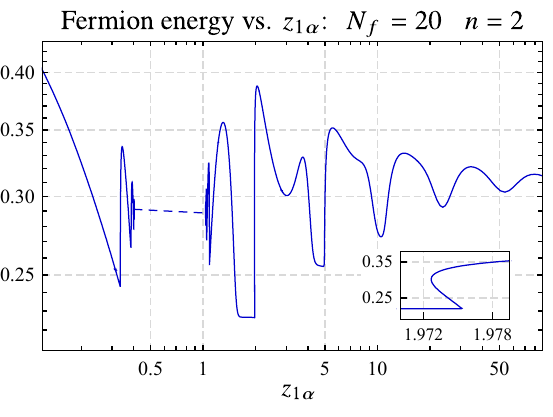}\hspace{5pt}
	\includegraphics{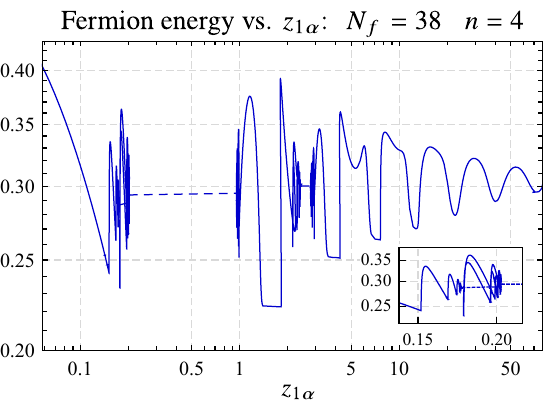}	
	\caption{The results of parametrizing the families of excited states by the redshift measured relative to the first fermion node within each solution. \textbf{Top row:} The $n=2$ fermion energy curves, plotted as a function of central redshift, for $N_f=12$ (left) and $N_f=20$ (middle), and the $n=4$ fermion energy curve for $N_f=38$ (right). \textbf{Bottom row:} The corresponding fermion energy curves plotted as a function of $z_{1\alpha}$. The new parametrization is largely successful, although multi-valued portions remain for $N_f=20$ and $N_f=38$ (see insets).}
	\label{figApp}
\end{figure*}

First, recall that the existence of the multi-valued regions (folds) in the fermion energy-redshift relations is due to the occurrence of solutions that contain fermion-field nodes within their power-law zones. As shown previously (see Fig.~\ref{figN20nodes}), these nodes transition smoothly from the wave zone to the power-law zone and back again along each fold. This suggests constructing `redshift-like' parameters that measure the value of the metric field $T(r)$ at the radius of each node, relative to its value at the origin. We therefore define
\begin{equation}
z_{p\alpha}=\frac{T(0)}{T(r_{p\alpha})}-1,
\end{equation}
where $r_{p\alpha}$ is the radius of node $p$ (counting outwards from the origin) in $\alpha(r)$. Since $T(r)$ decreases monotonically from its central maximum, this quantity is guaranteed to be positive. For solutions in which the node in question is located within the wave zone, $T(r_{p\alpha})$ will be close to unity, and so $z_{p\alpha}$ will only differ slightly from the central redshift (hereafter referred to as $z_0$). If the node is within the power-law zone, however, the difference will be significant. Of course, a similar parameter can also be defined based on $\beta(r)$, but since each node in $\alpha(r)$ is accompanied by a node in $\beta(r)$ at slightly larger radius, the two quantities will be almost identical.

The result of this new parametrization is illustrated in Fig.~\ref{figApp}. The leftmost plots show the fermion energy of the family of first even-parity excited states, for the case of $N_f=12$, plotted as a function of both central redshift $z_0$ (top) and $z_{1\alpha}$ (bottom). As can be seen, the parametrization by $z_{1\alpha}$ does indeed entirely remove the multi-valued portion of the curve, as desired.
	
For the case of $N_f=20$ (middle plots), however, it does not prove quite so successful. Firstly, there is a small portion of the curve around $z_{1\alpha}=1.97$ which remains multi-valued (see inset), corresponding to the region around the redshift transition point at $z_0=2.1$. We suspect that this could be removed by considering not only the node in $\alpha(r)$ but also the node in $\beta(r)$, although we have been unable to obtain the required combination. Secondly, there is a gap in the curve from $z_{1\alpha}\approx 0.4$ to $z_{1\alpha}\approx 1$ (the points on either side of which have been joined by a dashed line). This arises because the fold in the fermion energy-redshift relation extends beyond the upper redshift limit of our numerics, preventing us from obtaining all the solutions that lie along it. It is not clear, however, whether the entirety of this gap would in fact be bridged by including these high-redshift solutions. The rapid decay of the oscillations on either side of the gap suggests that this might not be the case. If so, then $z_{1\alpha}$ would no longer be a continuous parametrization of the curve.

The application of this method to the families of higher excited states is not so straightforward. For the first excited states, there is only a single node in $\alpha(r)$, and therefore the only quantity that can be constructed is $z_{1\alpha}$. For higher excited states, however, in which there exist multiple fermion nodes, we have a choice of parameters --- which one should we use? For low values of $N_f$, it turns out that $z_{1\alpha}$ remains the quantity that results in single-valued curves. This is because the first node in $\alpha(r)$ is the only one that transitions into the power-law zone; the others remain in the wave zone as the curve is traversed. Above $N_f\approx 24$, however, solutions appear in which multiple nodes exist in the power-law zone, and $z_{1\alpha}$ is no longer the appropriate parameter to use. This is illustrated in the rightmost plots of Fig.~\ref{figApp}, for the case of $N_f=38$, showing the fermion energy curve for the family of second even-parity excited states ($n=4$). The parametrization by $z_{1\alpha}$ appears largely successful, but zooms of the oscillatory regions reveal that there remain significant multi-valued portions (see inset). These regions correspond to the two folds that contain solutions with two nodes within their power-law zones (such as A, B, E, and F in Fig.~\ref{figN40n4}). Along these folds, the first fermion node remains relatively static, while it is the second that transitions from the power-law to the wave zone. These regions can therefore be made single-valued by using the parameter $z_{2\alpha}$. The trade-off, however, is that the remainder of the curve then becomes multi-valued once more.

Overall, it would theoretically be possible to obtain a single-valued fermion energy curve by switching between parameters depending on the structure of the solutions located along each fold. This is not entirely satisfactory, however, as it requires prior knowledge of the entire curve, and furthermore could not be used to assign a unique value to each excited-state solution. It may of course be possible to construct a single-valued quantity that involves a combination of the parameters $z_{p\alpha}$, although our attempts at doing so have been unsuccessful. Alternatively, it may be that a different approach is required.

\end{document}